\documentclass[fleqn,usenatbib]{mnras}

\usepackage{newtxtext,newtxmath}

\usepackage[T1]{fontenc}

\DeclareRobustCommand{\VAN}[3]{#2}
\let\VANthebibliography\thebibliography
\def\thebibliography{\DeclareRobustCommand{\VAN}[3]{##3}\VANthebibliography}

\usepackage{graphicx}	
\usepackage{amsmath}	
\usepackage{lscape}     

\newcommand{\kms}{km~s$^{-1}$}

\newcommand{\unitlum}{erg~s$^{-1}$}
\newcommand{\Msun}{M$_{\odot}$}

\title[The Tidal Disruption Event AT2022wtn]{The case of AT2022wtn: a Tidal Disruption Event in an interacting galaxy}

\author[F. Onori et al.]{
F. Onori,$^{1}$\thanks{E-mail: francesca.onori@inaf.it}
M. Nicholl,$^{2}$
P. Ramsden,$^{3,2}$
S. McGee,$^{3}$
R. Roy,$^{4}$
W. Li,$^{5}$
I. Arcavi,$^{5}$
J.~P. Anderson,$^{6,7}$
\newauthor
E. Brocato,$^{1,30}$
M. Bronikowski,$^{8}$
S.~B. Cenko,$^{9,10}$
K. Chambers,$^{11}$
T.~W. Chen,$^{12}$
P. Clark,$^{13}$
E. Concepcion,$^{8}$
\newauthor
J. Farah,$^{14,15}$
D. Flammini,$^{16}$
S. Gonz\'alez-Gait\'an,$^{17,6}$
M. Gromadzki,$^{18}$
C.~P. Guti\'errez,$^{19,20}$
E. Hammerstein,$^{21,9,22}$
\newauthor
K.~R. Hinds,$^{23}$
C. Inserra,$^{24}$
E. Kankare,$^{25}$
A. Kumar,$^{26}$
L. Makrygianni,$^{27}$
S. Mattila,$^{25,28}$
K.~K. Matilainen,$^{25}$
\newauthor
T.~E. Müller-Bravo,$^{20,19}$
T. Petrushevska,$^{8}$
G. Pignata,$^{29}$
S. Piranomonte,$^{30}$
T. M. Reynolds,$^{25,31,32}$
R. Stein,$^{33}$
\newauthor
Y. Wang,$^{34}$
T. Wevers,$^{35, 38}$
Y. Yao,$^{36,37}$
D.~R. Young.$^{2}$
\\
$^{1}$INAF-Osservatorio Astronomico d'Abruzzo, via M. Maggini snc, I-64100 Teramo, Italy\\
$^{2}$Astrophysics Research Centre, School of Mathematics and Physics, Queens University Belfast, Belfast BT7 1NN, UK\\
$^{3}$Institute of Gravitational Wave Astronomy \& School of Physics and Astronomy, University of Birmingham, Edgbaston, Birmingham, B15 2TT, UK\\
$^{4}$Manipal Centre for Natural Sciences, Manipal Academy of Higher Education, Karnataka, Manipal - 576104, India\\
$^{5}$School of Physics and Astronomy, Tel Aviv University, Tel Aviv 69978, Israel\\
$^{6}$European Southern Observatory, Alonso de C\'ordova 3107, Casilla 19, Santiago, Chile\\
$^{7}$Millennium Institute of Astrophysics MAS, Nuncio Monsenor Sotero Sanz 100, Off 104, Providencia, Santiago, Chile\\
$^{8}$Center for Astrophysics and Cosmology, University of Nova Gorica, Vipavska 11c, 5270 Ajdov\v{s}\v{c}ina, Slovenia\\
$^{9}$Astrophysics Science Division, NASA Goddard Space Flight Center, Greenbelt, MD 20771, USA\\
$^{10}$Joint Space-Science Institute, University of Maryland, College Park, MD 20742, USA\\
$^{11}$Department of Physics and Astronomy, University of Hawaii, 2680 Woodlawn Drive, Honolulu, HI 96822, USA\\
$^{12}$Graduate Institute of Astronomy, National Central University, 300 Jhongda Road, 32001 Jhongli, Taiwan\\
$^{13}$Institute of Cosmology and Gravitation, University of Portsmouth, Portsmouth, PO1 3FX, UK\\
$^{14}$Las Cumbres Observatory, 6740 Cortona Drive, Suite 102, Goleta, CA 93117-5575, USA\\
$^{15}$Department of Physics, University of California, Santa Barbara, CA 93106-9530, USA\\
$^{16}$ENEA, NUC Department, Via E. Fermi 45, 00044 Frascati, Rome, Italy\\
$^{17}$Instituto de Astrof\'isica e Ci\^encias do Espaço, Faculdade de Ci\^encias, Universidade de Lisboa, Ed. C8, Campo Grande, 1749-016 Lisbon, Portugal\\
$^{18}$Astronomical Observatory, University of Warsaw, Al. Ujazdowskie 4, 00-478 Warszawa, Poland\\
$^{19}$Institut d'Estudis Espacials de Catalunya (IEEC), Edifici RDIT, Campus UPC, 08860 Castelldefels (Barcelona), Spain \\
$^{20}$Institute of Space Sciences (ICE, CSIC), Campus UAB, Carrer de Can Magrans, s/n, E-08193 Barcelona, Spain \\
$^{21}$Department of Astronomy, University of Maryland, College Park, MD 20742, USA\\
$^{22}$Center for Research and Exploration in Space Science and Technology, NASA/GSFC, Greenbelt, MD 20771, USA\\
$^{23}$Astrophysics Research Institute, Liverpool John Moores University, Liverpool Science Park, 146 Brownlow Hill, Liverpool L3 5RF, UK\\ 
$^{24}$Cardiff Hub for Astrophysics Research and Technology, School of Physics \& Astronomy, Cardiff University, Queens Buildings, The Parade, Cardiff, CF24 3AA, UK\\
$^{25}$Department of Physics and Astronomy, University of Turku, FI-20014 Turku, Finland\\
$^{26}$Department of Physics, University of Warwick, Coventry CV4 7AL, UK\\
$^{27}$Department of Physics, Lancaster University, Lancaster LA1 4YB, UK\\
$^{28}$School of Sciences, European University Cyprus, Diogenes Street, Engomi, 1516, Nicosia, Cyprus\\
$^{29}$Instituto de Alta Investigación, Universidad de Tarapacá, Casilla 7D, Arica, Chile\\
$^{30}$INAF, Osservatorio Astronomico di Roma, Via di Frascati 33, I-00078 Monteporzio Catone, Italy\\
$^{31}$Cosmic Dawn Center (DAWN), Niels Bohr Building, Radmandsgade 62-64 2200, Copenhagen, Denmark\\
$^{32}$Niels Bohr Institute, University of Copenhagen, Jagtvej 128, DK-2200, Copenhagen N, Denmark\\
$^{33}$Division of Physics, Mathematics, and Astronomy, California Institute of Technology, Pasadena, CA 91125, USA\\
$^{34}$Key Laboratory of Optical Astronomy, National Astronomical Observatories, Chinese Academy of Sciences, Beijing 100101, People’s Republic of China\\
$^{35}$Space Telescope Science Institute, 3700 San Martin Drive, Baltimore, MD 21218, USA\\
$^{36}$Miller Institute for Basic Research in Science, 468 Donner Lab, Berkeley, CA 94720, USA\\
$^{37}$Department of Astronomy, University of California, Berkeley, CA 94720, USA\\
$^{38}$Astrophysics \& Space Institute, Schmidt Sciences, New York, NY 10011, USA
}

\date{Accepted XXX. Received YYY; in original form ZZZ}

\pubyear{2024}

\begin{document}
\label{firstpage}
\pagerange{\pageref{firstpage}--\pageref{lastpage}}
\maketitle
\clearpage

\begin{abstract}
We present the results from our multi-wavelength monitoring campaign of the transient AT\,2022wtn, discovered by the Zwicky Transient Facility in the nucleus of SDSS\,J232323.79+104107.7, the less massive galaxy in an active merging pair with a mass ratio of $\sim$10:1. 
AT\,2022wtn shows spectroscopic and photometric properties consistent with a X-ray faint N-strong TDE-H+He with a number of peculiarities. Specifically, a 30-days long plateau at maximum luminosity, a corresponding dip in temperature and the development of a double-horned \ion{N}{III}+\ion{He}{II} line profile. Strong and time-evolving velocity offsets in the TDE broad emission lines and the detection of a transient radio emission, indicate the presence of outflows. Overall, the observed properties are consistent with the full disruption of a low-mass star by a $\sim$10$^{6}$\Msun\/ SMBH followed by an efficient disk formation and the launch of a quasi-spherical reprocessing envelope of fast expanding outflowing material. 
The observed differences between the \ion{He}{II} and the Hydrogen and \ion{N}{III} lines can be explained either with a spatial separation of the lines emitting region or with a late-time reveal of shocks from the returning debris streams, as the photosphere recedes. Finally, we present an extensive analysis of the hosting environment and discuss the implications for the discovery of two TDEs in interacting galaxy pairs, finding indication for an over-representation of TDEs in these systems. The AT\,2022wtn host galaxy properties suggest that it is in the early stages of the merger, therefore we may be witnessing the initial enhanced rate of TDEs in interacting galaxies before the post-starburst phase.
\end{abstract}




\section{Introduction}
When a star approaches a supermassive black hole (SMBH), it can be ripped apart by the strong tidal forces at play if they overcome the star's self-gravity. This results in a tidal disruption event \cite[TDE,][]{rees88, phinney89,evans89}. During these phenomena, approximately half of the stellar material is expelled in unbound orbits, while the rest streams back to the SMBH, it is stretched into elongated streams, and it starts a circularization process into highly eccentric orbits, which ends with the formation of a new accretion disk around the involved SMBH. The peak of the TDE emission is in the X-rays and/or UV/optical bands, and in some cases it can even exceed the Eddington luminosity of the SMBH \cite[][]{lodato11,strubbe09, wevers19}. Observationally, TDEs give rise to a short-lived and luminous (L$_{\rm bol} \sim$10$^{41-45}$ \unitlum) transient flare located in the nuclear regions of the hosting galaxy. These galaxies are typically quiescent, with an observed overabundance of TDEs detected in post-starburst/E+A host galaxies \cite[][]{arcavi14, french16, graur18, french20}. Interestingly, a recent work of \citet{wevers24} has shown that TDEs are strongly overrepresented in gas-rich post-starburst galaxies with faded AGNs and extended emission line regions, indicating that they may have had a merger very recently.

These phenomena are particularly interesting for a number of reasons. They represent an extraordinary laboratory for studying black holes (i.e. they are used to directly constrain the mass and spin of Black Holes (BHs)) and accretion-related phenomena on human-friendly time scales. Given that the star's disruption can only occur outside the BH horizon event, they are an excellent tool to unveil dormant SMBHs in the low-mass end of the SMBH mass distribution (M$_{BH} \leq$10$^{7}$\Msun, above $\sim$10$^{8}$ \Msun\/ the star is swallowed whole), including the elusive population of intermediate-mass BHs. 
The occurrence of TDEs can be used to probe the SMBH occupation fraction in different types of galaxies \cite[][]{metzgerstone16}, and their rate and their evolution over time can provide important information on the existence of SMBHs at high redshift \cite[][]{mortlock11}. Finally, TDEs are multi-messenger phenomena, being candidate sources of high-energy neutrinos \cite[][]{stein21,reusch22}, and gravitational wave (GW) sources potentially detectable by future space-based interferometers \cite[][but see also the work of \citet{wevers23} for different results on the TDEs GW detectability]{toscani20, ajith24}.  

TDEs were predicted to be brightest in X-rays, and indeed the first candidates were discovered as luminous X-ray transient sources in the {\it ROSAT} all sky survey archive \cite[][]{komossa99} and, subsequently, other TDEs were revealed through dedicated searches or serendipitous discoveries with the {\it Chandra}, {\it XMM-Newton} and {\it Swift} satellites \cite[][]{esquej07,esquej08, saxton12, komossa15, saxton20}. However, in the last decade, thanks to the availability of increasingly efficient wide-field optical surveys, specifically designed to search for transient phenomena, the sample of TDEs has rapidly grown from a few candidates to tens of confirmed TDEs and the optical band has become their primary discovery channel. A population of transients with some well established key observational features has been revealed, but also characterized by a broad range of properties, with each event still providing new clues but raising new questions \cite[see the reviews from][]{vanVelzen20, gezari21}. Indeed, recent researches found a few nuclear transients with ambiguous properties, which were initially classified as TDEs (cf. \citealt{2020MNRAS.494.2538N, 2022ApJ...930...12H, 2024MNRAS.528.6176R} and references therein).

The UV/optical light-curves are characterized by persistent blue colors (i.e. g-r$<$0), relatively long timescales for the rise to the peak luminosity ($<$30 days) if compared to most typical supernovae (SNe), and a smooth, power-law decline broadly consistent with $L\propto$t$^{-5/3}$ law (following the predicted fall-back rate on to the
BH), which can last from months to years \cite[although recently a great diversity in light curve shapes has been revealed,][]{vanvelzen11, vanVelzen20}. The blackbody (BB) temperatures are usually high (T$_{\rm BB} \sim$10$^{4}$ K) and approximately constant over the whole transient evolution, but they are also accompanied by a time-evolving BB radius \cite[][]{vanVelzen21}.   
In the early phases, the optical spectra are typically dominated by a strong, hot and blue thermal continuum and by very broad ($\sim$10$^{4}$ km s$^{-1}$) H and/or He lines characterized by a pure emission profile and with different relative ratios \cite[][]{arcavi14, leloudas19, charalampopoulos22}. In some cases, broad Bowen fluorescence emission lines and evidence for the presence of \ion{Fe}{II} multiplets have been identified \cite[][]{blagorodnova19, leloudas19, onori19, wevers19, cannizzaro21}. This spectral variety led to the division of the TDE population into three main spectral classes, on the basis of the presence or lack of the main broad emission lines: the TDE-H, TDE-H+He (most of them showing also Bowen features) and TDE-He \cite[][]{arcavi14, vanVelzen20, vanVelzen21}. However, this classification may be time dependent, as evidenced by some TDEs with spectral lines that appear or disappear in time \cite[e.g. AT2017eqx, AT\,2017gge][]{nicholl19, onori22}.
The interaction of the TDE flare with the host galaxy’s environment can result in reverberation signals \cite[IR echoes and transient long-lasting high ionization coronal emission lines,][]{lu16, vanvelzen16,jiang21,onori22,short23}, which trace the circumnuclear hosting environment and can be used as additional discovery channels. Indeed, recently, a number of TDEs have been identified also thanks to the observation of transient mid-infrared flares \cite[][]{mattila18,kool20,jiang21,reynolds22,masterson24}.

Thanks to the growth of the TDE sample together with the increasing use of a multiwavelength approach in the observing strategies, a crucial dichotomy in the TDE population has been revealed. Specifically, among the optically selected TDEs, only few events show X-rays at the time of the optical peak, and instead a delayed X-ray emission was detected in some notable cases \cite[e.g ASASSN-14li, ASASSN-15oi, AT\,2019dsg, AT\,2018fyk, AT\,2019qiz, AT\,2019azh, AT\,2017gge;][respectively]{holoien16b,gezari17,cannizzaro21,wevers19b,nicholl20,liu22,onori22}. Interestingly, from recent studies it has emerged that a large fraction of optical-selected X-rays faint TDEs ($>$40\%) do show X-rays emission at later times \cite[][]{jonker20,guolo24}.
These discoveries have led to questions about both the origin of the TDE prompt emission mechanism and the properties of the emission region itself and different theoretical scenarios have been thus proposed, which are still under debate. The strong UV/optical emission is ascribed either to the occurrence of shocks between self-intersecting debris streams well before the end of the circularization process and the formation of the accretion disk \cite[][]{piran15, shiokawa15, jiang16,steinberg24}, or to reprocessing of the X-rays emitted by a newly formed accretion disk \cite[][]{guillochon13, guillochon14,metzgerstone16, roth16}. The latter scenario needs the presence of an optically thick reprocessing-layer which can arise either as a consequence of super-Eddington accretion induced outflows \cite[][]{miller15, thomsen22} or collision-induced outflows at the streams self-intersection point \cite[][]{lu_bonnerot20, bonnerot21} or a promptly and rapidly circularization of the debris stream which cools inefficiently \cite[cooling envelope model,][]{metzger22}. 
In the TDE unified model of \cite{dai18}, optically thick winds produced following the formation of the accretion disk are responsible for the X-ray obscuration and viewing angle effects together with the intrinsic physical properties of the electromagnetic emitting region determine the detection of the X-ray emission. 

\subsection{AT\,2022wtn: the discovery}
The discovery of AT\,2022wtn was first announced by the Zwicky Transient Facility \cite[ZTF,][]{bellm19} on 2022 October 2 (MJD 59\,854.26), which reported the observation of a new optical transient at magnitude $g^\prime$ = 19.57 mag, internally labelled as ZTF22abkfhua. The transient is located at coordinates RA (J2000) = 23:23:23.772 DEC (J2000) = +10:41:07.83, consistent with the nucleus of the galaxy SDSS J232323.79+104107.7. This galaxy is undergoing a merger with the more massive galaxy SDSS J232323.37+104101.7, as shown in the Legacy Survey DR10 image of the AT\,2022wtn field (see Figure \ref{fig:legacy}). Thanks to spectroscopic observations on 2022 November 21, the nuclear transient was identified as a TDE at z=0.049 \cite[TNS Classification Report No. 13873,][]{fulton22}. 

AT\,2022wtn was observed  with NSF's Karl G. Jansky Very Large Array (VLA) on 2023 January 8 and 2023 March 21 (MJD 59\,952 and MJD 60\,024, respectively) at a mean frequency of 15 GHz and the detection of an emission coincident with the AT\,2022wtn position was reported by \cite{christy23} (ATel $\#$15972). In particular, while in the first epoch the detection was only marginal, the source clearly brightened in the second observation, rising from a flux density of 20$\pm$7$\micron$Jy to a flux density of 223$\pm$6 $\micron$Jy and, thus, confirming the transient nature of the detected radio emission.  

We here report the result from the UV/optical and X-ray follow-up campaign covering a total of $\sim$393 days from the transient discovery. Throughout the paper all the phases are referenced with respect to the transient's discovery date (unless otherwise specified) and a luminosity distance of $d_{\rm L}$ = 219.9 Mpc, based on a WMAP9 cosmology with $H_{\rm 0}$ = 69.32 km s$^{-1}$ Mpc$^{-1}$ , $\Omega_{\rm M}$ = 0.29, $\Omega_{\Lambda}$ = 0.71 \cite[][]{hinshaw2013} is used.

\section{Observations and data reduction}
\label{sec:observations} 

In order to provide a good coverage and characterization of the AT\,2022wtn emission, we promptly started a long-lasting monitoring campaign, characterized by a multiwavelength approach and including photometric and spectroscopic observations. 
%
In the following, we describe the observational set-up and the data reduction for each instrument used.

\subsection{Optical photometry}
\label{subsec:phot}

\begin{figure}
\centering
\includegraphics[width=\columnwidth]{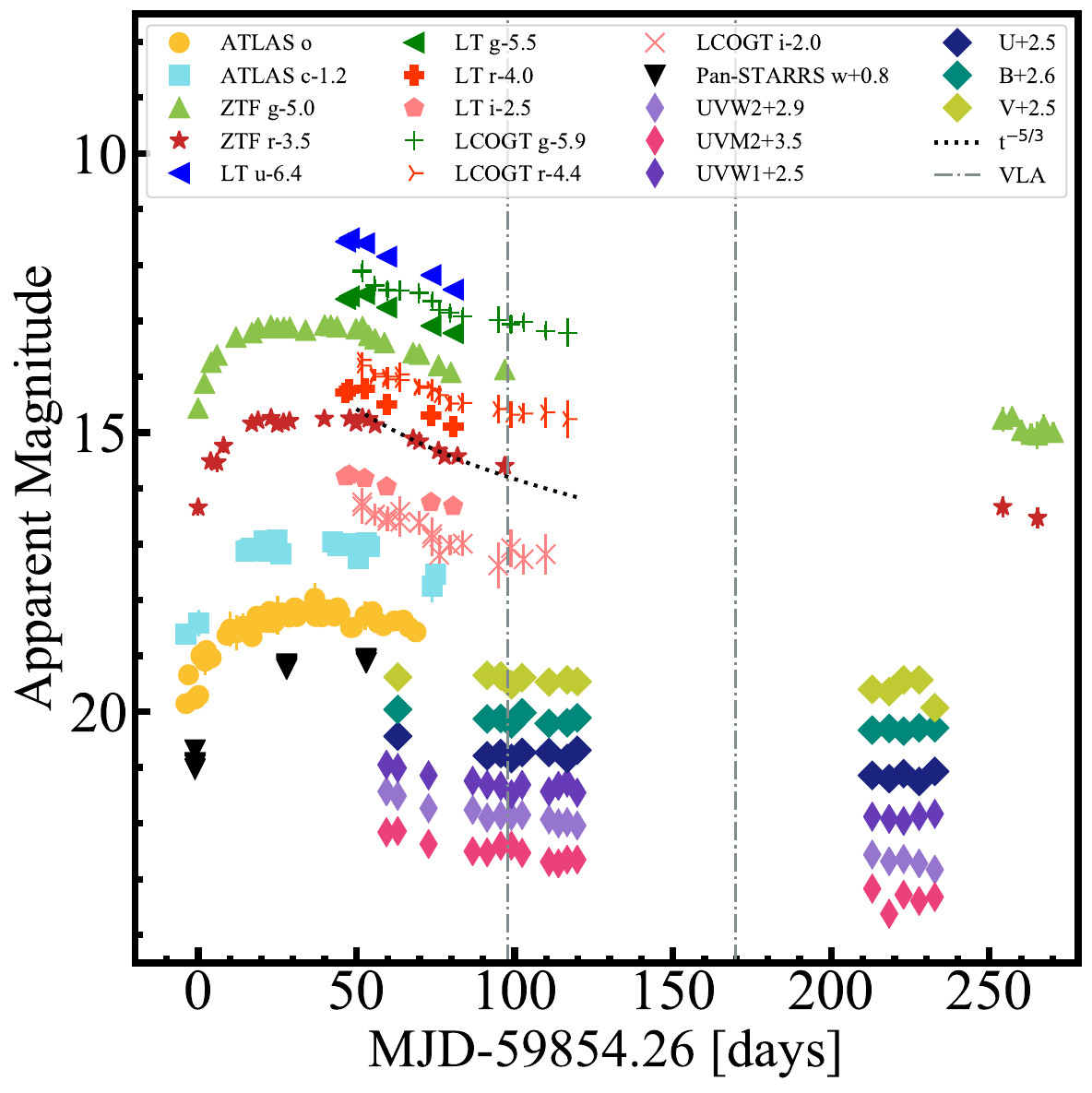}
\caption{AT\,2022wtn UV/optical light-curve. The optical data are host-subtracted. Magnitudes are reported in the AB system and are not corrected for foreground extinction. Vertical dot-dashed grey lines show the location of the VLA detection. We show the t$^{-5/3}$ decline with a dotted line overplotted on the ZTF $r^{\prime}$ points. 
}
\label{fig:lc}
\end{figure}

In Figure \ref{fig:lc} the AT2022wtn UV/optical light-curve is shown for a total of 250 days. The rise and the peak phases (corresponding to the first $\sim$ 100 days) are densely monitored by the ZTF, the Asteroid Terrestrial-impact Last Alert System \cite[ATLAS,][]{smith20} and and the Panoramic Survey Telescope and Rapid Response System \cite[Pan-STARRS,][]{huber15, chambers16} transient surveys, which delivered host-subtracted magnitudes measurements in the $g^\prime$, $r^\prime$, $c$, $o$ filters, respectively. ZTF data have been retrieved from Lasair alert stream broker \cite[][]{smith19}; calibrated ATLAS data in the cyan ($c$) and orange ($o$) bands were obtained using the ATLAS forced photometry service \citep{tonry18,smith20,Shingles2021}. We rebinned the four exposures obtained on each night into a single nightly measurement to improve the signal-to-noise ratio. Pan-STARRS host-subtracted magnitudes in the $w$ band are also available during the first $\sim$50 days of the transient evolution.
Pan-STARRS observations are processed and photometrically calibrated with the PS image processing pipeline \cite[][]{Magnier2020,Magnier2020a,Waters2020}.

We obtained additional images in the $u^\prime$, $g^\prime$, $r^\prime$ and $i^\prime$ filters with the Optical Wide Field Camera (IO:O), mounted on the 2.0m  Liverpool Telescope \cite[LT,][]{steele04} located at the Roques de los Muchachos Observatory in La Palma island (Spain), and with the Las Cumbres Observatory \cite[LCOGT\footnote{https://lco.global},][]{brown13} global network of 1.0m telescopes.  These observations cover the post-peak phase between $\sim$50 and $\sim$100 days from the transient discovery. 
We processed the IO:O/LT images by using the instrument pipeline\footnote{https://telescope.livjm.ac.uk/TelInst/Pipelines/\#ioo}, while the photometry was derived by using a custom pipeline, Photometry Sans Frustration\footnote{https://github.com/mnicholl/photometry-sans-frustration} \citep{nicholl23}, which is a \texttt{PYTHON}-based code employing both aperture and point-spread-function (PSF) fitting photometry routines from \texttt{ASTROPY} and \texttt{PHOTUTILS}. The $g^\prime$, $r^\prime$ and $i^\prime$ zero points were   derived by using reference stars from Pan-STARRS, while for the $u^\prime$ we used SDSS stars. Image subtraction was performed by using \texttt{PyZOGY} \cite[][]{zackay16,guevel17} and pre-transient templates images from Pan-STARRS ($g^\prime$, $r^\prime$ and $i^\prime$) and SDSS ($u^\prime$). Finally, the photometry measurements were performed by fitting the PSF on each host-subtracted images. 

Also in the case of the optical imaging data from LCOGT
the suppression of the host-galaxy contamination have been performed by using the \texttt{PyZOGY} package, while the photometric measurements have been performed by applying PSF photometry with the AUTOmated Photometry Of Transients pipeline \cite[\texttt{AUTOPHOT},][]{brennan22}. In Table \ref{tbl:Optphot} we list the photometric measurements derived for the host-subtracted I:IO and LCOGT data. The apparent magnitudes are reported in the AB system and not corrected for foreground extinction. Instead, archival photometry available for both for the AT\,2022wtn host and the neighbouring galaxy is reported in Table \ref{tab:phot_host}. 

\subsection{UVOT photometry and X-ray Non-detection}
We accurately monitor the UV and X-ray emission with the the UVOT and XRT instruments on board of the Neil Gehrels {\it Swift} observatory \cite[][]{gehrels04} following the transient peak. 
Specifically, a total of 17 {\it Swift}/UVOT+XRT observations have ensured the monitoring of the UV/optical and X-rays emission between $\sim$60 and $\sim$230 days from the transient discovery (see Table \ref{tbl:UVOTphot} for the log of the observations). 

The {\it Swift}/UVOT observations include images in the filters \textit{UVW2} (1928 \AA), \textit{UVM2} (2244 \AA), \textit {UVW1} (2600 \AA), \textit {U} (3465 \AA), \textit {B} (4392 \AA) and \textit {V} (5468 \AA) which have been reduced using the standard pipeline with the updated calibrations from the \texttt{HEAsoft}-6.28 \texttt{ftools} package. 
In order to derive the apparent magnitudes of the transient we used the \texttt{HEAsoft} routine \texttt{uvotsource}. For each filter we measured the aperture photometry using a 5\arcsec\/ aperture centered on the position of the transient and  a background region of 60\arcsec\/ radius placed in an area free of sources. In Table \ref{tbl:UVOTphot} we show the values obtained from the {\it Swift}/UVOT photometric measurements for all the filters. The apparent magnitudes are reported in the Vega system and uncorrected for foreground extinction. In the last row of Table \ref{tbl:UVOTphot}, we also report the host contribution in the UVOT filters as derived from the host galaxy spectral energy distribution (SED) fitting, described in more detail in Section \ref{subsec:sed}. We used these synthetic host photometric values to subtract the host galaxy contribution in all the UVOT filters but the B and V bands, where the galaxy is much brighter. For these two cases, we measured the host galaxy flux using the last UVOT epoch (MJD 60\,087), when the host galaxy dominates the transient contribution. 

Along with the UVOT, the {\it Swift} onboard XRT telescope also observed the transient in the energy range 0.3$-$10 keV. However, nothing was detected in XRT, even after stacking all of
the X-ray images of the field (count rate $<$ 1.89$\times10^{-3}$ counts\,sec$^{-1}$).
The X-ray non-detection of AT2022wtn along with the detection of some of the probable 
Bowen fluorescence lines may have significant implications in understanding the nature 
of the transient (see \S\ref{sec:bowen}).




\begin{figure}
\centering
\includegraphics[width=0.8\columnwidth]{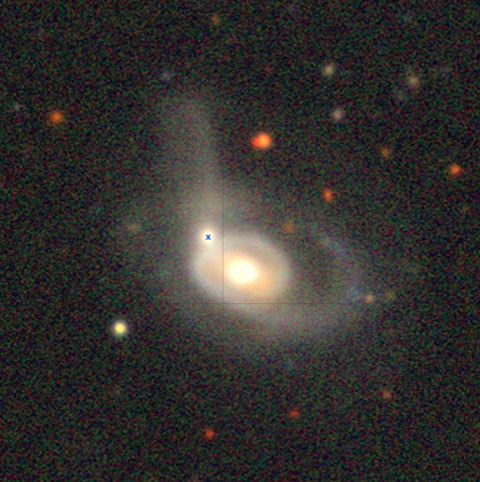}
\caption{Legacy Survey DR10 image of the AT\,2022wtn field. The TDE occurred on the nucleus of the smaller interacting galaxy as indicated by the blue cross. The tidal tails resulting from the merging interaction between the two galaxies are well visible.}
\label{fig:legacy}
\end{figure}

\begin{table}
\centering
\caption{Archival photometry of the AT\,2022wtn host and for the interacting galaxy, SDSS J232323.37+104101.7, taken from 
SDSS DR16 ($u^\prime$; $g^\prime$; $r^\prime$; $i^\prime$; $z^\prime$, in AB system), 
and from 2MASS ($J$; $H$; $K_s$) catalogs.}
\begin{tabular}{lcc}
\hline
Filters	 & SDSS J232323.79+104107.7 & SDSS J232323.37+104101.7 \\
(mag)    & (AT\,2022wtn host)       &                           \\
\hline
u$^{\prime}$ & 18.987$\pm$0.084&17.75$\pm$0.02\\
g$^{\prime}$ & 17.305$\pm$0.008&16.03$\pm$0.00\\
r$^{\prime}$ & 16.692$\pm$0.007&15.21$\pm$0.00\\
i$^{\prime}$ & 16.301$\pm$0.008&14.75$\pm$0.00\\
z$^{\prime}$ & 16.062$\pm$0.019&14.42$\pm$0.00\\
J            & 15.950$\pm$0.150&14.125$\pm$0.052\\
H            & 15.451$\pm$0.221&13.363$\pm$0.063\\
K$_S$        & 14.790$\pm$0.192&12.932$\pm$0.057\\
\hline
\end{tabular}
\label{tab:phot_host}
\end{table}

\subsection{Optical spectroscopy}
\label{subsec:spec}
Our spectroscopic follow-up started $\sim$5 days after the transient discovery and it has been carried out by using the following facilities:  the SED Machine \cite[SEDM][]{blagorodnova18} mounted on the Palomar 60-inch telescope (P60), located at the Palomar Observatory (USA), the SPectrograph for the Rapid Acquisition of Transients \cite[SPRAT,][]{piascik14} mounted on the LT, the FLOYDS spectrograph installed on the LCOGT 2m telescope Faulkes Telescope North (FTN) located at the Haleakala observatory (Hawaii), the DeVeny Spectrograph mounted at the 4.3-meter Lowell Discovery Telescope (LDT) located in Flagstaff (Arizona), the ESO Faint Object Spectrograph and Camera \cite[EFOSC2,][]{buzzoni84}, mounted on the 3.58 m New Technology Telescope (NTT) and the Low Resolution Imaging Spectrometer \cite[LRIS,][]{oke95} operating at the Cassegrain focus of the 10-meter Keck I telescope, located at the W. M. Keck Observatory in Maunakea island (Hawaii). Two spectra of the neighbour galaxy SDSS J232323.37+104101.7 have been also taken with the the DeVeny and the EFOSC2 spectrographs. 

All the spectra were reduced in a standard manner which include bias, flat-field, cosmic ray correction and wavelength and flux calibration via arc lamp and standard star spectra.  
In Table \ref{tab:spec_log} the list of the spectroscopic observation and their main properties, such as the date of the observation, the instrument used, the exposure times, the airmass and seeing, is reported. The sequence of the optical spectra is shown in Figure \ref{fig:spec}.

\subsection{The spectroscopic instrumentation}
In the following, we describe the details of the spectroscopic instrumentation used:

\begin{itemize}
\item The first spectrum of AT\,2022wtn was taken 4.98 days after the transient discovery with SEDM. This is a very low resolution spectrograph operating at $R=\lambda/\Delta\lambda \sim 100$ in the 4000 - 8000 \AA\/ wavelength range. Data are reduced by using the SEDM automatic pipeline\footnote{https://github.com/MickaelRigault/pysedm}.
\item We obtain two AT\,2022wtn spectra with SPRAT at 23.66 days and 49.58 days from the transient discovery. This low resolution (R$\sim$350, for a 1.8$\arcsec$ slit width) spectrograph, operates in the 4020 - 8100 \AA\/ wavelength range. Data are reduced by using an adaption of the \texttt{frodospec}\footnote{https://telescope.livjm.ac.uk/TelInst/Pipelines/\#frodospecL2} automatic pipeline. 
\item Two additional spectra at 53.20 and 62.17 days were observed with the FLOYDS spectrograph on the FTN of the LCOGT network. These spectra cover 3500 to 10,000 \AA, at a resolution $R \approx 400-700$ (at the blue and red end, respectively) and were taken at the paralactic angle. 
The spectra were reduced using the \texttt{floydsspec} pipeline\footnote{\url{https://github.com/svalenti/FLOYDS_pipeline/}} and the final spectrum extraction is described in \cite{Valenti2014}. 
\item The DeVeny spectrograph works in a broad wavelength range, between 3200 \AA\/ and 1 $\mu$m, with a moderate resolution ranging between R=500 and R=4000, depending upon the grating used. We obtain one AT\,2022wtn spectrum at 59.91 days and one spectrum of the neighbour interacting galaxy SDSS J232323.37+104101.7 during the same night. We used the grating DV2 which covers the 3000 - 7400 \AA\/ wavelength range with a dispersion of 2.17 \AA\//px and R=920. 
Data have been reduced by using the python based pipeline \texttt{PypeIt}\footnote{https://pypeit.readthedocs.io/en/latest/}. 
\item We have obtained a total of three AT\,2022wtn spectra and one spectrum of the neighbour galaxy (SDSS J232323.37+104101.7) with the EFOSC2 instrument in the framework of the advanced ePESSTO+ ESO public spectroscopic survey \cite[][]{smartt15}. In all the cases we used the grism Gr$\#$13, which covers the 3650–9250 \AA\/ wavelength range and provides a resolution of  $R=\lambda/\Delta\lambda \sim 300$ for a 1\arcsec.0 slit (a slightly higher resolution is achieved for seeing better than 1\arcsec.0). We used a slit width of 1\arcsec.0 or 1\arcsec.5, depending on the seeing condition during the observations, oriented at the parallactic angle. All the spectra have been reduced by using the \texttt{ePESSTO NTT Pipeline v.2.4.0}\footnote{https://github.com/svalenti/pessto}, which is based on standard IRAF tasks.
Multiple spectra taken on the same night are averaged in order to increase the SNR.
\item One spectrum was taken with LRIS 105.96 days. This instrument is characterized by a broad wavelength coverage (3200 - 10000 \AA\/), obtained thank to the simultaneous observation of two arms (operating in the red and in the blue part of the spectrum, respectively). A variety of available gratings (red side) and grisms (blue side) yield resolutions ranging from R=300 and R=5000. Our observations were carried out in long slit mode, with a slit width of 1$\arcsec$.0 oriented at the parallactic angle. We used the the grism 400/3400 for the blue camera, yielding a R$\sim$600 (calculated for a 1$\arcsec$.0 slit width) and the grating 400/8500 for the red camera, which yields to a resolution of R$\sim$1000 (for a 1$\arcsec$.0 slit width). Data are reduced with the instrument dedicated software package \texttt{LPipe} \cite[][]{perley19}.
\end{itemize}











\begin{table*}
\centering
\caption{List spectroscopic observation of AT\,2022wtn. (1) Date of observation; (2) days after discovery; (3) Instrument used; (4) spectral resolution; (5) Exposure time; (6) Airmass; (7) Seeing.\\ 
$^{*}$ spectrum of the neighbour galaxy SDSS J232323.37+104101.7}
\begin{tabular}{lllcccc}
\hline
MJD			& phase & Instrument & R & Exp. Time & Airmass & Seeing \\
(days)      & (days)&            &($\uplambda/\Delta\uplambda$) &(s)        & (arcsec)& (arcsec) \\
(1)         & (2)   & (3)        & (4)&(5)        & (6)     & (7)       \\
\hline
59\,859.24  & \phantom{10}4.98 & SEDM    & 100 & 2250          &1.086 &\\
59\,877.91  & \phantom{1}23.66 & SPRAT-B & 350 & \phantom{1}600 & 1.056 &2.256\\
59\,903.84  & \phantom{1}49.58 & SPRAT-B & 350 & 2100 & 1.057 &1.068\\
59\,907.46  & \phantom{1}53.20 & FLOYDS  & 400-700 & 3600 & 1.711 &2.785\\
59\,914.18  & \phantom{1}59.91 & DeVeny  & 920  & 900   & 1.25 &1.00   \\ 
59\,914.18$^{*}$  & \phantom{1}59.91 & DeVeny  & 920 &  900   & 1.25 &1.00   \\
59\,916.43  & \phantom{1}62.17 & FLOYDS     & 400-700 & 3600 & 1.639 & 2.729 \\ 
59\,934.04  & \phantom{1}79.78 & EFOSC2-Gr\#13 & 300 & 2700 & 1.778 & 0.970\\ 
59\,960.22  & 105.96 & KECK-LRIS & 600/1000 & \phantom{1}600 & 1.532 & \\
60\,086.39  & 232.13 & EFOSC2-Gr\#13 & 300 & 2$\times$1800 & 1.581 & 2.30 \\ 
60\,086.39$^{*}$ & 232.13 & EFOSC2-Gr\#13 &300 & 2$\times$1800 & 1.581 & 2.30 \\ 
60\,247.10  & 392.84 & EFOSC2-Gr\#13 & 300 & 2$\times$1800 & 1.330 & 1.10 \\ 
\hline
\end{tabular}
\label{tab:spec_log}
\end{table*}

\section{Photometric analysis}
\label{sec:superbol}
In Figure \ref{fig:lc} we show the AT\,2022wtn optical/UV light-curve, with magnitude offsets applied to each filter for clarity. All the optical magnitudes are host-subtracted, but not yet corrected for foreground extinction.  Already at this initial stage of the analysis, the overall light-curve behaviour is clear. It is characterized by a peak value of $r^\prime$=18.34$\pm$0.05 mag, reached after a rising phase of $\sim$20 days. Remarkably, it remains around this value for $\sim$30 days before starting the declining phase. 

In order to investigate on the physical properties of the UV/optical emission, we computed the bolometric luminosity by using the AT\,2022wtn host-subtracted multi-color photometry as input values in the \texttt{python}-based routine \texttt{superbol} \cite[][]{nicholl18}. All the used magnitudes are corrected for the Galactic extinction from \cite{schlafly11} assuming a reddening law with $R_{V}=3.1$ and Av=0.2073 (mag), K-correction \cite[][]{oke68} has been also applied. When needed, we derived the missing magnitudes values by extrapolating the photometry assuming a constant color evolution for the light curve. We calculated the pseudo-bolometric luminosity by integrating over the observed fluxes and by fitting with a single BB function the SED inferred from the multi-color data for each epoch. We used the best-fitted BB model to compute the additional flux bluewards of the UVW2 band and redwards of $i$ band. 

\begin{figure}
\includegraphics[width=\columnwidth]{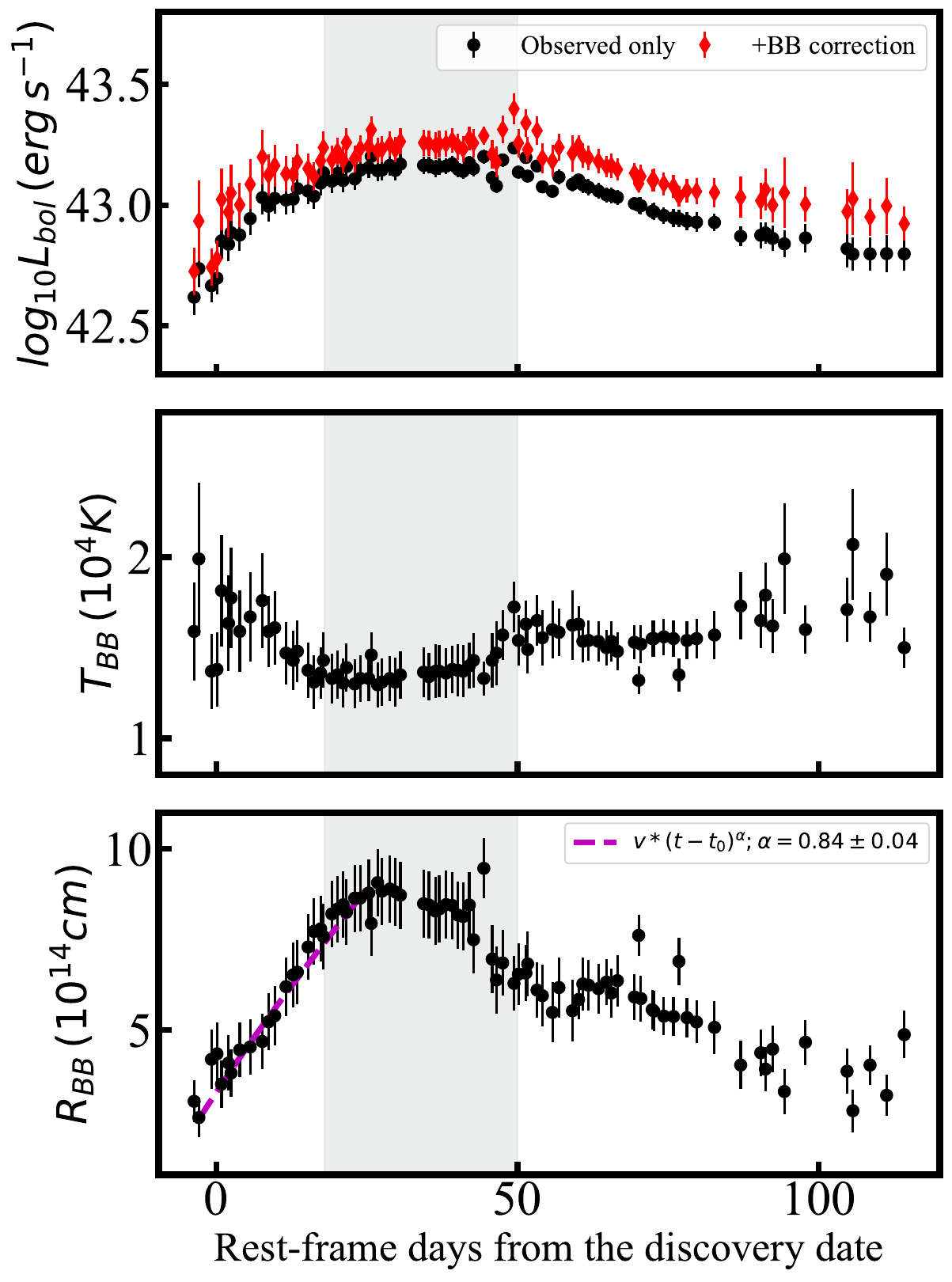}
\caption{\texttt{Superbol} output showing the pseudo-bolometric lightcurve (upper panel), the $T_{\rm BB}$ (central panel) and the $R_{\rm BB}$ (bottom panel) time evolution. The input magnitudes are host-subtracted and have been corrected for reddening. The data are plotted with a bin of 1 day. Grey area indicate the duration of the maximum luminosity plateau phase. The magenta dashed line in the bottom panel shows the best fit for the $R_{\rm BB}$ rising phase.}
\label{fig:bolometric}
\end{figure}

\begin{figure}
\includegraphics[width=\columnwidth]{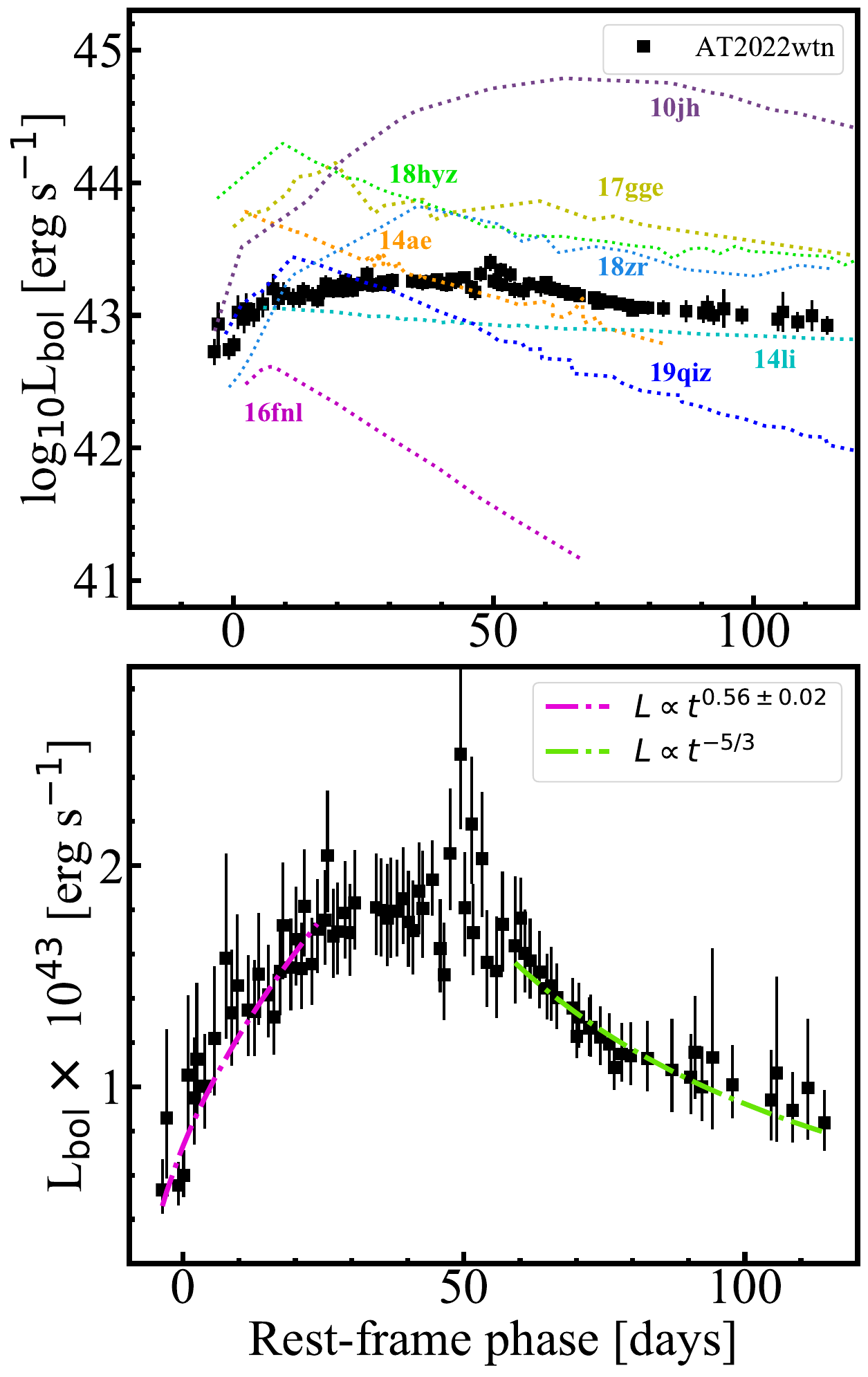}
\caption{Upper panel: pseudo-bolometric luminosity of AT2022wtn in comparison with other TDE lightcurves. Lower panel: results from the fit of the rising (magenta dashed-dotted line) and declining (green dashed-dotted line) phase of the AT\,2022wtn pseudo-bolometric luminosity.}
\label{fig:lbol}
\end{figure}

The results are shown for the first 100 days of the transient evolution in Figure \ref{fig:bolometric}. In the upper panel, the pseudo-bolometric luminosity derived by using the two different methods is shown. The BB temperature and radius evolution with time are shown in the center and bottom panels, respectively. The same time evolution exhibited by the optical/UV lightcurve is clearly visible also in the pseudo-bolometric luminosity. Specifically, after an initial rising phase, it reaches a maximum luminosity plateau, lasting about 30 days, during which the luminosity has an averaged value of $<$log(L$_{\rm BB, max}$)$>$=43.13$\pm$0.06 erg s$^{-1}$. By using the \texttt{python} tool \texttt{curve$\_$fit}, we modelled the rise of the pseudo-bolometric luminosity with a power-law of the form $L_{\rm BB}$=L$_{(\rm BB,0)}$ ((t-t$_{0}$)/$\tau$)$^{a}$ and we obtain an initial time t$_{0}$=-6.6$\pm$-2.6 days, a rising timescale $\tau$=11.5$\pm$3.4 days and a power-law index a=0.56$\pm$0.14. Moreover, we also model the first 50 days of the decline of the lightcurve with the powerlaw $L_{\rm BB}$ $\propto$((t-t$_{0}$)/$\tau$)$^{-5/3}$, by fixing t$_{0}$=-6.6$\pm$-2.6 days (i.e. the value obtained from the fit of the rising phase). The results are shown in the bottom panel of Figure \ref{fig:lbol}.

A comparison with the pseudo-bolometric light-curves of a compilation of TDEs, computed by applying the same method used for AT\,2022wtn is shown in the upper panel of Figure \ref{fig:lbol}. In particular, the photometric data of iPTF16fnl and of AT\,2017gge are taken from \cite{onori19} and \cite{onori22}, respectively, while, for the other TDEs we retrieve the data from \cite{gezari12, holoien14,holoien16a,vanvelzen19, gomez20, nicholl20}, respectively. 
AT\,2022wtn is placed in the middle of the others TDEs light-curves, with the luminosity consistent with the values usually expected for these transients. 

As expected in the case of a TDE, the BB temperature is consistent with being constant at a value of $T_{\rm BB} \sim$1.55$\times$10$^{4}$K, although, between $\sim$20 and $\sim$40 days from the transient discovery (consistent with the luminosity maximum plateau phase), it constantly remains at a slightly lower value ($T_{\rm BB} \sim$1.35$\times$10$^{4}$K). 
Instead, there is a clear evolution of the BB radius with time, starting from initial values of about $R{_{\rm BB}}\sim$3$\times$10$^{14}$cm and rising over the first 25 days of the transient evolution to a maximum value of about $R{_{\rm BB}}\sim$9$\times$10$^{14}$cm. The BB radius remains around this value for a total of about 20 days before starting a declining trend toward the initial value, reached again after $\sim$100 days of the transient evolution. The rising phase is well fitted by the function $R_{\rm BB}$=v (t-t$_{0}$)$^{\alpha}$ as shown with a magenta dashed line in the bottom panel of Figure \ref{fig:bolometric}. In particular we fixed the initial time to t$_{0}$=-6.6$\pm$-2.6 days, which is the value obtained from the pseudo-bolometric luminosity, and we obtain power-law with index $\alpha$=0.84$\pm$0.04 and an expansion velocity v=4974$\pm$578 km s$^{-1}$. It is interesting to note that the expansion phase of the BB radius last more days with respect to the rise of the luminosity, with the luminosity reaching the peak plateau well before the the reaching of the maximum value of the BB radius.     

The photometric properties shown so far, such as the powerlaw rise to the peak luminosity in about 20 days, the powerlaw decay well resembling the t$^{-5/3}$ behaviour, the constant BB temperature at $T_{\rm BB} \sim$1.55$\times$10$^{4}$ K but with an evolving BB radius, are all commonly observed in TDEs \cite[][and reference therein]{hinkle20, vanVelzen20,vanVelzen21,zabludoff21}, and thus they support the TDE nature of AT\,2022wtn.

\section{The spectroscopic features}
\label{sec:linefit}
In the following, we describe the spectroscopic analysis performed on the reduced spectra of AT\,2022wtn with the aim to investigate on the transient's emission lines properties and evolution.
The reduced spectra have been first corrected for foreground extinction by using the Cardelli law \cite[][]{Cardelli89} with E(B-V)=0.0651 mag  and assuming a reddening law with R$_{V}$=3.1 \cite[][]{schlafly11}. 
Given the lack of an archival spectrum of the host galaxy and the presence of broad features related to the transient emission in the latest spectra we have, the analyzed spectra are not host-subtracted. However, we have removed the contribution of the transient blue continuum by modelling it with a 3rd order \texttt{spline3} with the \texttt{iraf} task \texttt{continuum}.
We remark that, as emerged in the late-time spectra, the host galaxy has strong star-forming emission lines which, although narrow, could affect or even mask the presence of faint broad TDE lines, particularly in the low S/N spectra.

\subsection{The emission line fit}\label{sec:emissionline}
The emission lines detected in the continuum-corrected spectra are modelled by using a Gaussian function with the \texttt{python} fitting packages \texttt{curvefit} and \texttt{leastsq}. In the case of line profiles characterized by the simultaneous presence of more features (i.e. both broad and narrow components), a multi-component Gaussian model has been used. During the fitting procedure a wavelength window of $\sim$1000 \AA\/ around the the spectral features of interest has been selected in order to include also the local continuum. The properties of the different components, such as the central wavelength ($\uplambda_{c}$) and the full width at half maximum (FWHM), have been left as free parameters. In Figure \ref{fig:line_fit} we show the results (fit and residuals with respect to the model) for the \ion{He}{II}+H$\upbeta$ and the \ion{He}{I}+H$\upalpha$ regions of the spectra taken at 23.66 days, 49.58 days, 59.91 days, 79.78 days, 105.96 days and 232.13 days from the transient discovery. In Table \ref{tab:line_fit} we report the fitting parameters ($\uplambda_{c}$, FWHM and the equivalent width (EW)) derived for the \ion{N}{III}, Hydrogen and Helium emission lines for each spectrum.

\subsubsection{The spectroscopic evolution}
\label{sec:bowen}

The first spectrum, taken with SEDM 4.98 days after the transient discovery, shows only a blue continuum and some narrow emission lines, such as the H$\beta$, [\ion{O}{III}]$\uplambda$5007, H$\upalpha$ and the [\ion{S}{II}] $\uplambda \uplambda$6716,6730  doublet blend which we ascribe to the host galaxy contribution. No broad components are clearly detected at this stage, however, we note that this is a low S/N spectrum, thus we could probably be not sensitive to the potential presence of broad but weak features.

A spectral evolution is seen in the subsequent sequence of spectra. Specifically, the spectra taken with LT/SPRAT at 23.66 days and 49.58 days from the transient discovery \cite[the first one it has been used for the spectroscopic classification,][]{fulton22} show both broad and narrow emission lines (Figure \ref{fig:spec}). In particular, broad components with FWHM$\sim$10$^{4}$ \kms\/
are found in correspondence of \ion{N}{III}$\lambda$4640 and H$\upalpha$, respectively, while a faint but well centered broad component (FWHM$\sim$6600 \kms\/ and FWHM$\sim$1.5$\times$10$^{4}$ \kms, at 23.66 days and 49.58 days, respectively) is detected at a position consistent with the \ion{He}{I}$\uplambda$5875. A faint H$\beta$ appears at 23.66 days which intensify in the subsequent spectrum, where it is detected with an intermediate width of FWHM$\sim$4$\times$10$^{3}$ \kms. We also note the emergence of a new emission line centered around $\uplambda_{c}$=4503.46 \AA\/ in the 49.58 days spectrum, which we tentatively identify with the \ion{N}{III}$\uplambda$4510. Instead, we do not detect any plausible feature in correspondence of the \ion{He}{II}$\uplambda$4686. We highlight that discriminating a broad \ion{N}{III}$\lambda$4640 from a blueshifted \ion{He}{II}$\uplambda$4686 can be a difficult process due to their wavelength proximity \cite[as shown in][]{nicholl19} and requires spectroscopic observations with a resolution higher than that obtained with SPRAT.
However, the fact that the broad \ion{N}{III}$\lambda$4640 is well centered at the expected wavelength ($\uplambda_c$=4641.77$\pm$7.27 \AA, see Table \ref{tab:line_fit}) and the evolution seen in the following spectra support this identification, although it is not conclusive in excluding any possible contribution of a \ion{He}{II} component to the overall line profile.   

Starting from 59.91 days from the transient discovery, we detect the emergence of a complex spectral feature composed by a series of different components in the \ion{He}{II} $\uplambda$4686+H$\upbeta$ region (see the left panels of Figure \ref{fig:line_fit}). We clearly detect a broad feature (FWHM$\sim$6$\times$10$^{3}$ \kms) at 59.91 days, still ascribable to the \ion{N}{III}$\uplambda$4640, although with a redshifted central wavelength ($\uplambda_c$=4667.60$\pm$5.19 \AA).
Furthermore, we observe the development of a clear \ion{He}{II}$\uplambda$4686 component as a double-horne shape in the \ion{N}{III}$\uplambda$4640 line profile. We note that this feature intensifies with time but it never develops a very broad component, keeping its width nearly constant at FWHM$\sim$4$\times$10$^{3}$ \kms\/ for all the time it is detected.  


The evolution of the FWHM and of the EW of all the broad and intermediate components detected in the AT\,2022wtn spectral sequence are shown in the upper and bottom panels of Figure \ref{fig:fwhm_evolution}, respectively. Specifically, we clearly detect broad components in the \ion{N}{III}$\uplambda$$\uplambda$4100,4640 all showing an evolution with time of the FWHM, although reaching different values of the maximum width and with different timing (see the upper-left panel of Figure \ref{fig:fwhm_evolution}). In particular, both the \ion{N}{III}$\uplambda$$\uplambda$4100,4640 are still well detected at later times (393 days from the transient discovery) with FWHM$\sim$7$\times$10$^{3}$\kms and FWHM$\sim$1.6$\times$10$^{4}$\kms, respectively. Instead, an intermediate-width \ion{He}{II}$\uplambda$4686 component is detected at later phases, after the luminosity peak, it never shows component exceeding the FWHM$\sim$4$\times$10$^{3}$\kms\/ and it is a long lasting feature (still detected in the last spectrum we have, taken at 393 days, upper-left panel of Figure \ref{fig:fwhm_evolution}). 

textbf{In the upper-right panel of Figure \ref{fig:fwhm_evolution}} we show the FWHM evolution with time for the Helium and Hydrogen emission lines. Specifically, we detect broad/intermediate components in the H$\upbeta$, \ion{He}{I}$\uplambda$5875 and H$\upalpha$. Differently to what observed in the \ion{He}{II}$\uplambda$4686, these others emission lines clearly show an evolution with time of their width. In particular, the \ion{He}{I}$\uplambda$5875 and the H$\upbeta$ are characterized by a widening trend, reaching maximum value for the line width of FWHM$\sim$2$\times$10$^{4}$\kms\/ before disappearing, while the H$\alpha$ shows narrowing trend. In any case, the broad components of all these lines are not detected anymore after $\sim$100 days from the transient discovery. 

The EW time evolution for the \ion{N}{III}$\uplambda$$\uplambda$4100,4640 and for the \ion{He}{II}$\uplambda$4686 are shown in the bottom-left panel of Figure \ref{fig:fwhm_evolution}. We note that, while the \ion{N}{III}$\uplambda$4100 feature show almost no time evolution of its EW, having a value of around 10 \AA\/ all the time, the EW of the \ion{N}{III}$\uplambda$4640 shows an initial declining trend, starting from $\sim$30 \AA\/ and reaching the value of $\sim$10 \AA\/ only at 59.91 days, when \ion{He}{II}$\uplambda$4686 is first detected. After this phase no EW evolution is seen also for these two features. A possible explanation for the initial high values of the \ion{N}{III}$\uplambda$4640 EW could reside in the blending of a faint and undetected \ion{He}{II}$\uplambda$4686 component.


The identification of broad components of FWHM$\sim$10$^{4}$ \kms\/ in the \ion{N}{III} and in the Hydrogen lines together with the observed spectroscopic evolution, are all compatible within the N-strong TDE scenario. 
The early detection of broad Bowen lines is a strong indication for the presence of a hidden accretion-related X-ray emission already ongoing, given that both a large flux of photons with $\uplambda <$228 \AA\/ and large optical depth in the obscuring/reprocessing photosphere (providing a multiple scattering regime) are needed in order to efficiently trigger the Bowen fluorescence mechanism, as outlined in \citet{leloudas19}. In this scenario, the broadening of the Bowen lines is primarily due to the scattering process, their width depends on the system viewing angle \cite[the wider the line the more edge on the system is observed,][]{dai18} and a narrowing trend with time of the FWHM is consistent with the decreasing of the optical depth of the line emitting region \cite[][]{roth18}. Thus, in the case of AT\,2022wtn, the non-detection of X-ray emission from the XRT monitoring coupled with the identification of very broad \ion{N}{III} lines in the early spectra are consistent with an accretion powered X-ray-faint TDE viewed closer to the disk.  


\subsubsection{The presence of outflows}
\label{sec:outflows}
Beside the detection and the observed evolution of the broad features described before, we also detect a strong evolution in the measured central wavelengths for some of them together 
with the presence of two very broad components in the \ion{He}{II}$\uplambda$4686+H$\beta$ region, which can be explained with the presence of an outflow.

\begin{figure}
\includegraphics[width=\columnwidth]{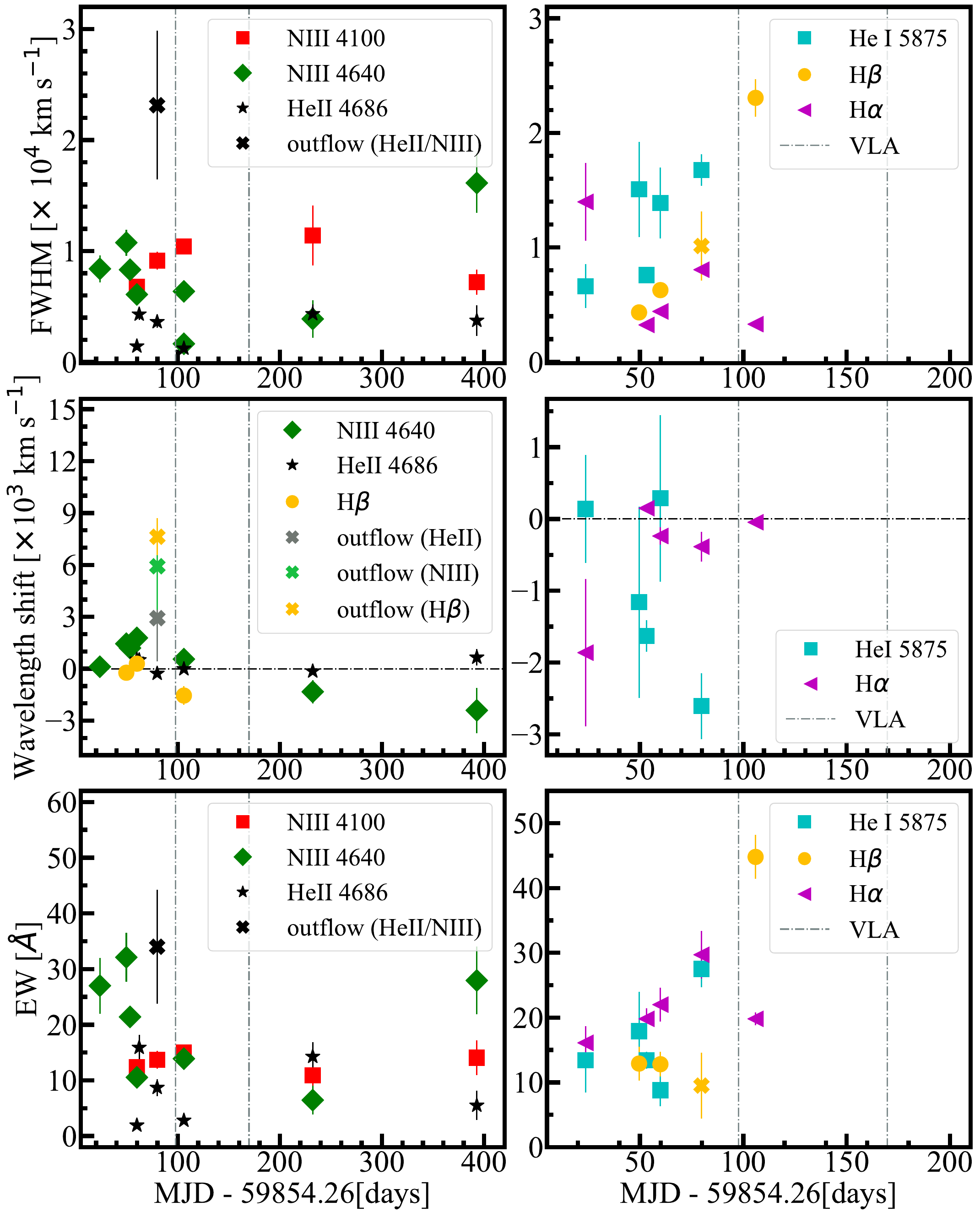}
\caption{{\it Upper panels:} FWHM evolution of the broad components detected in the AT\,2022wtn spectral sequence for the [\ion{N}{III}]$\uplambda\uplambda$4100;4640 and the \ion{He}{II}$\uplambda$4686; the H$\upbeta$, \ion{He}{I}$\uplambda$5875; H$\upalpha$ (left and right panels respectively). {\it Middle panels}: evolution of the central wavelength shift for the broad and intermediate components detected in the AT\,2022wtn spectral sequence for the [\ion{N}{III}]$\uplambda$4640, \ion{He}{II}$\uplambda$4686, H$\upbeta$, \ion{He}{I}$\uplambda$5875; H$\upalpha$ (left and right panels respectively). {\it Bottom panels}: EW evolution of the broad components detected in the AT\,2022wtn spectral sequence for the [\ion{N}{III}]$\uplambda\uplambda$4100;4640 and the \ion{He}{II}$\uplambda$4686; H$\upbeta$, \ion{He}{I}$\uplambda$5875; H$\upalpha$ (left and right panels respectively).
The colored cross markers indicate the possible outflows components. The dashed grey vertical line shows the location of the VLA detection.}
\label{fig:fwhm_evolution}
\end{figure}

In the central panels of Figure \ref{fig:fwhm_evolution}, we show the evolution of the wavelength shifts with respect to the expected wavelength value for the broad/intermediate components detected in the \ion{N}{III}$\uplambda$4640, \ion{He}{II}$\uplambda$4686, H$\upbeta$, \ion{He}{I}$\uplambda$5875 and H$\upalpha$ emission lines. In particular, the broad \ion{N}{III}$\uplambda$4640 is first detected with a central wavelength consistent with the restframe value ($\uplambda_{c}\sim$4641 \AA), however it subsequently shows a gradual wavelength shift which first moves toward the red, reaching a maximum velocity shift of v$\sim$2$\times$10$^{3}$ \kms\/ measured at $\sim$60 days, and then it shifts back toward the blue wavelengths, reaching a final velocity shift of v$\sim$-2$\times$10$^{3}$ \kms\/ at 293 days. 
The intermediate component in the \ion{He}{II}$\uplambda$4686 instead is always detected with a central wavelength consistent with its expected rest-frame value. Strong wavelength shift are also detected both in the H$\upalpha$ and in the \ion{He}{I}$\uplambda$5875 broad components. Specifically, the broad H$\upalpha$ is detected at early times with a blue-shifted central wavelength, corresponding to a velocity shift of v$\sim$-2$\times$10$^{3}$ \kms\/ and, as it narrows, it progressively come back the the expected central wavelength values. On the contrary, the \ion{He}{I}$\uplambda$5875 broad component is first detected at zero wavelength shift and, while broadening, it progressively show a blueshift, reaching a final velocity shift of v$\sim$-3$\times$10$^{3}$ \kms\/ at 79.78 days, before disappearing.
These lines velocity offsets identified in AT\,2022wtn are particular interesting as they can be used to probe the kinematics of the line forming region. Similar features in the emission lines have been already detected in a number of TDEs and, in some cases, have been attributed to the presence of outflows \cite[][]{nicholl20, charalampopoulos22}. Moreover, a time evolution in the velocity shift is predicted in the framework of the \citet{roth18} model on the electron scattering effects in shaping the profiles of the lines emitted in an hot and outflowing reprocessing photosphere. Thus, the properties observed in the AT\,2022wtn emission lines are consistent with being emitted in an outflowing reprocessing photosphere surrounding the SMBH.

\begin{figure}
\centering
\includegraphics[width=0.46\columnwidth]{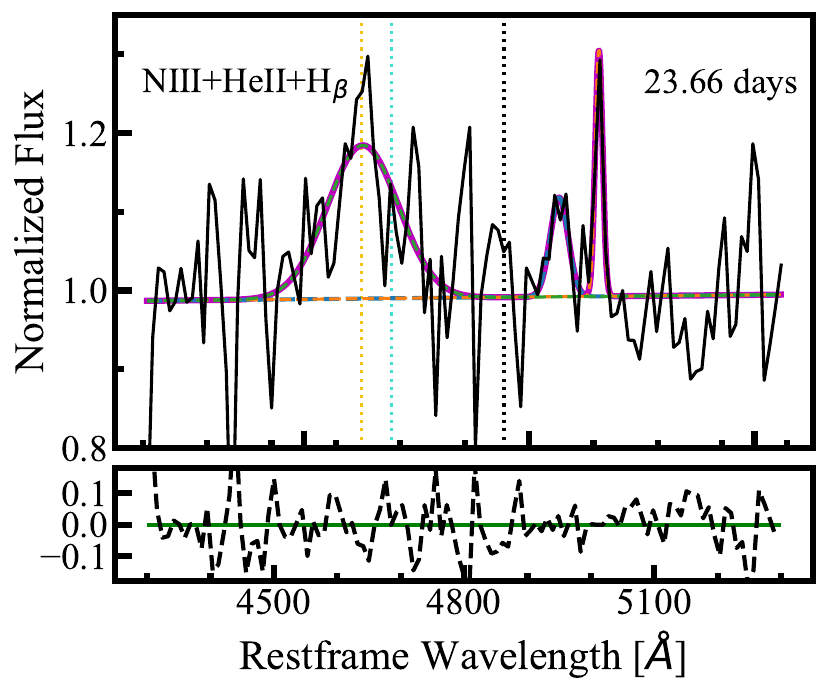}
\includegraphics[width=0.46\columnwidth]{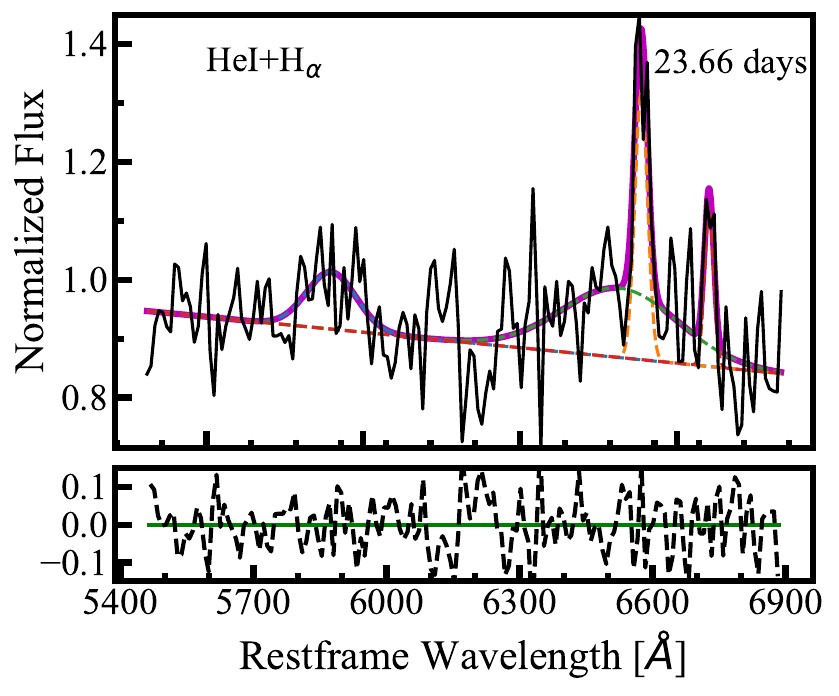}
\includegraphics[width=0.46\columnwidth]{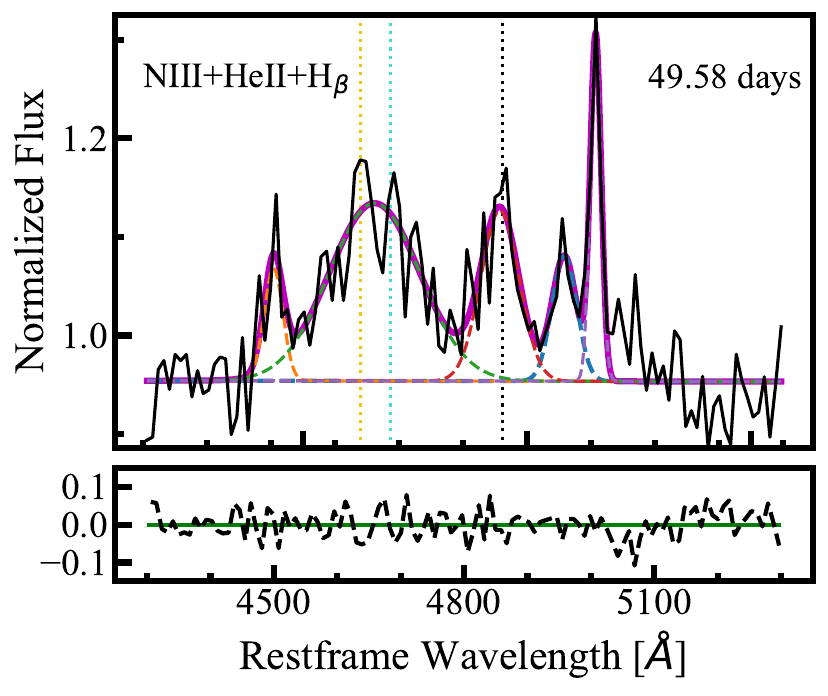}
\includegraphics[width=0.46\columnwidth]{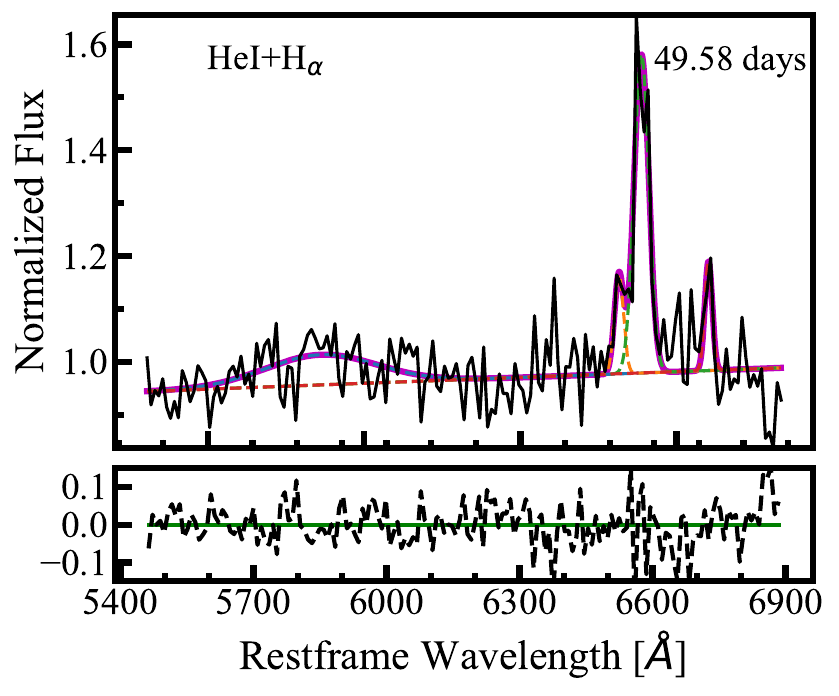}
\includegraphics[width=0.46\columnwidth]{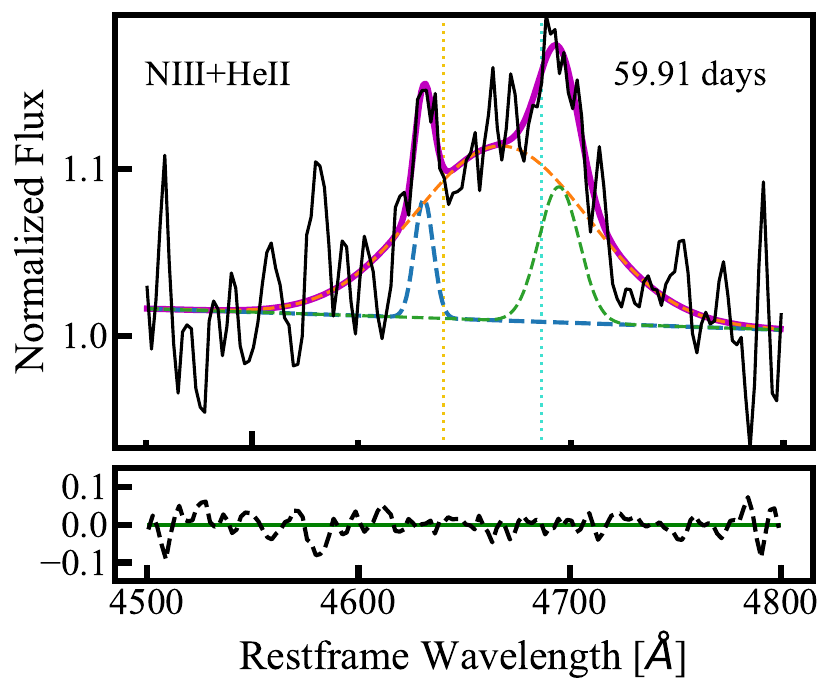}
\includegraphics[width=0.46\columnwidth]{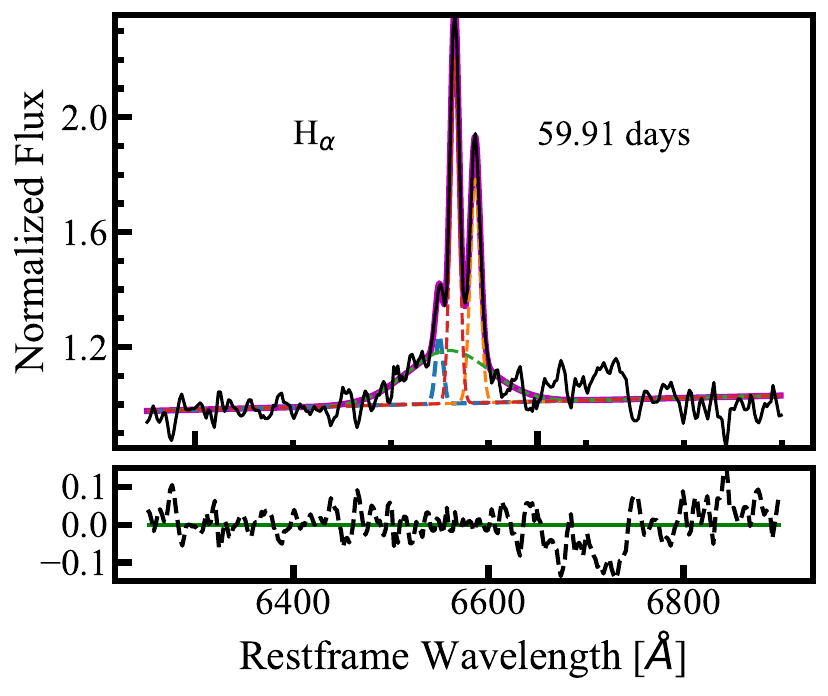}
\includegraphics[width=0.46\columnwidth]{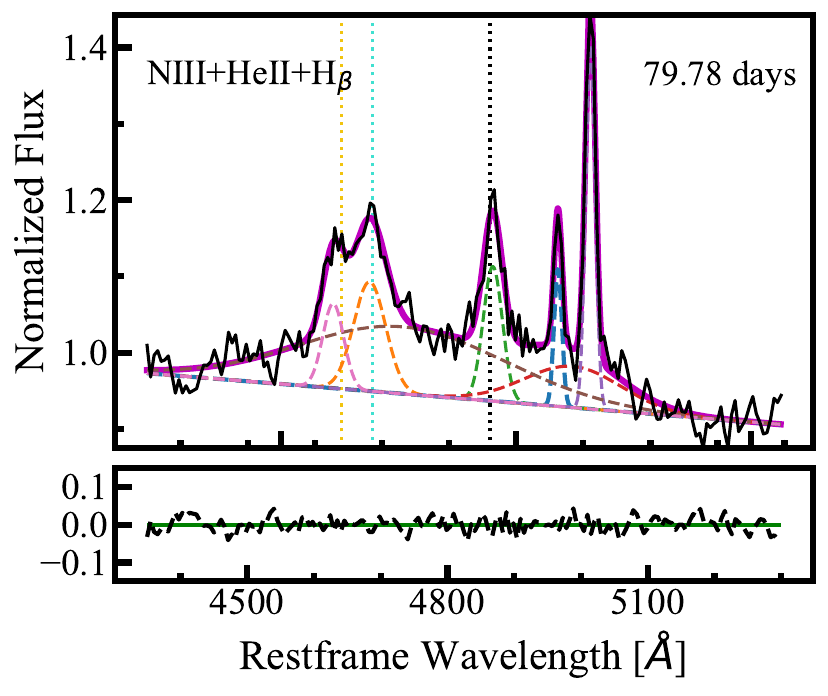}
\includegraphics[width=0.46\columnwidth]{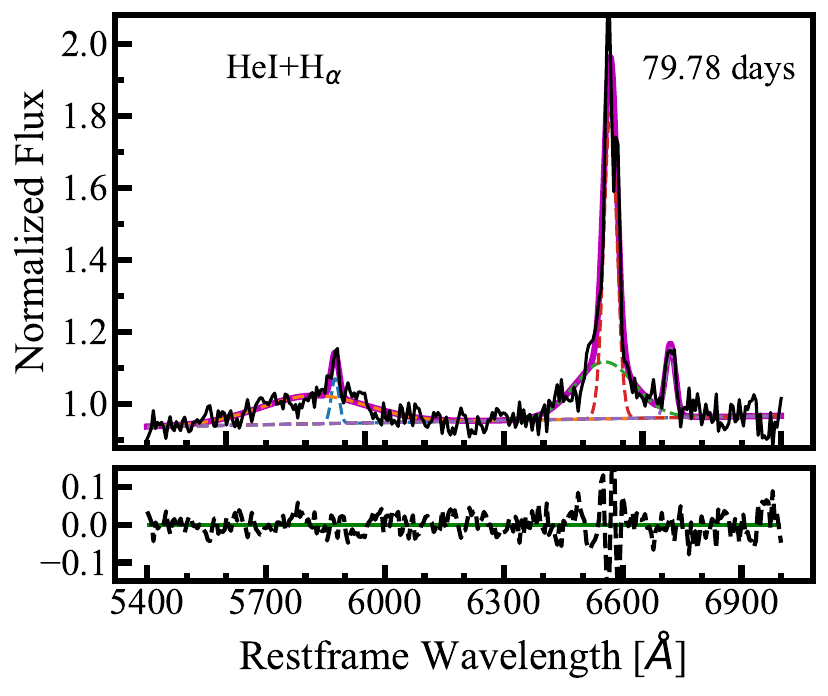}
\includegraphics[width=0.46\columnwidth]{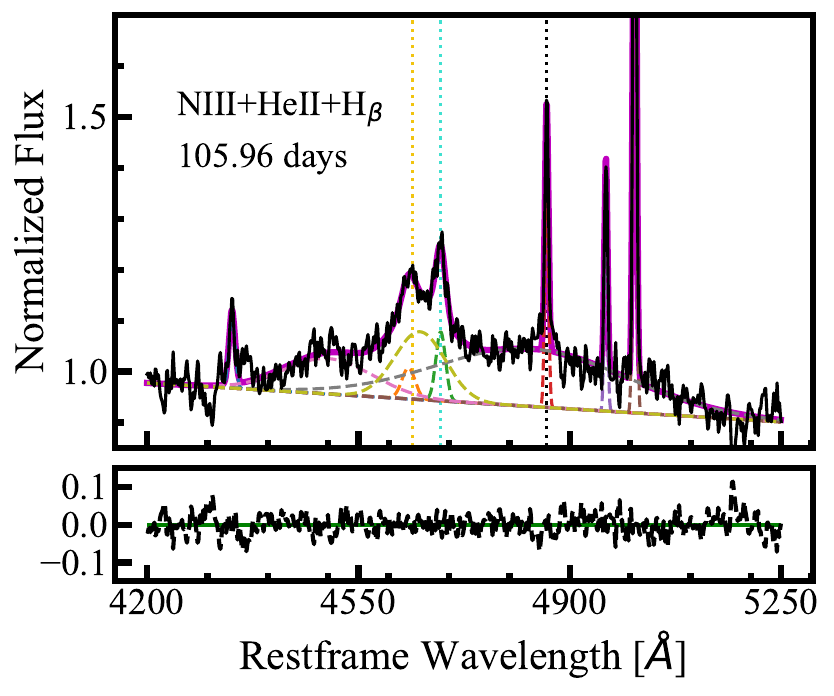}
\includegraphics[width=0.46\columnwidth]{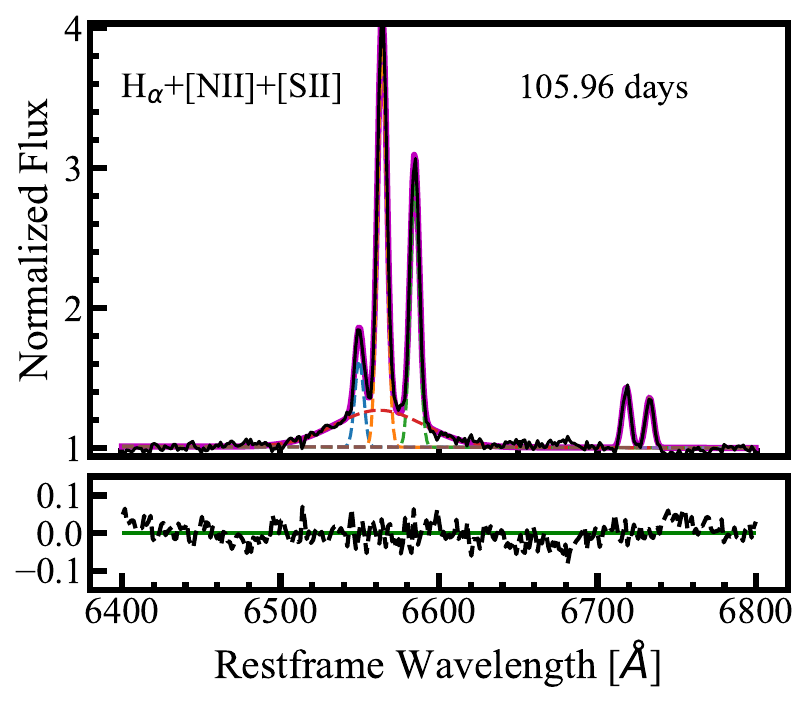}
\includegraphics[width=0.46\columnwidth]{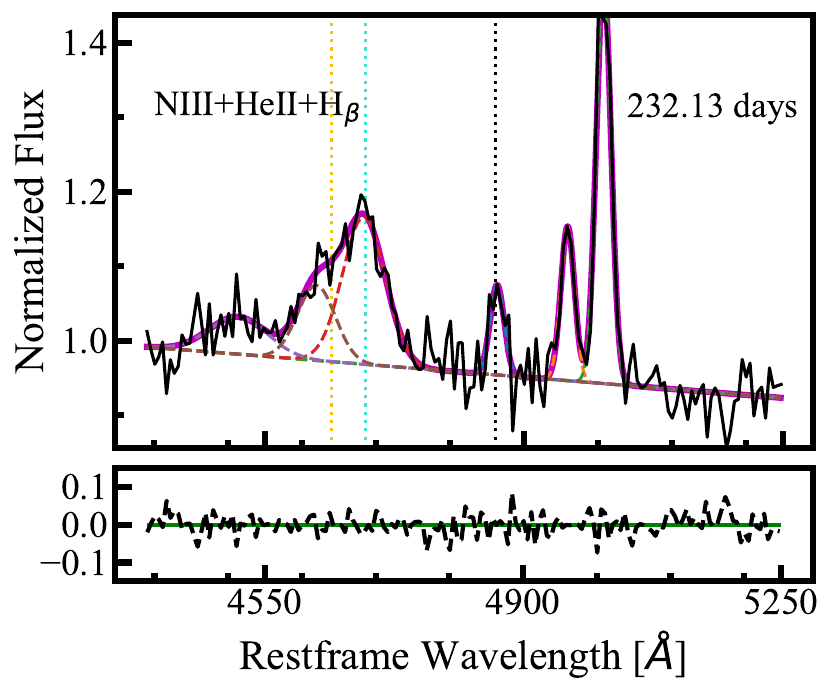}
\includegraphics[width=0.46\columnwidth]{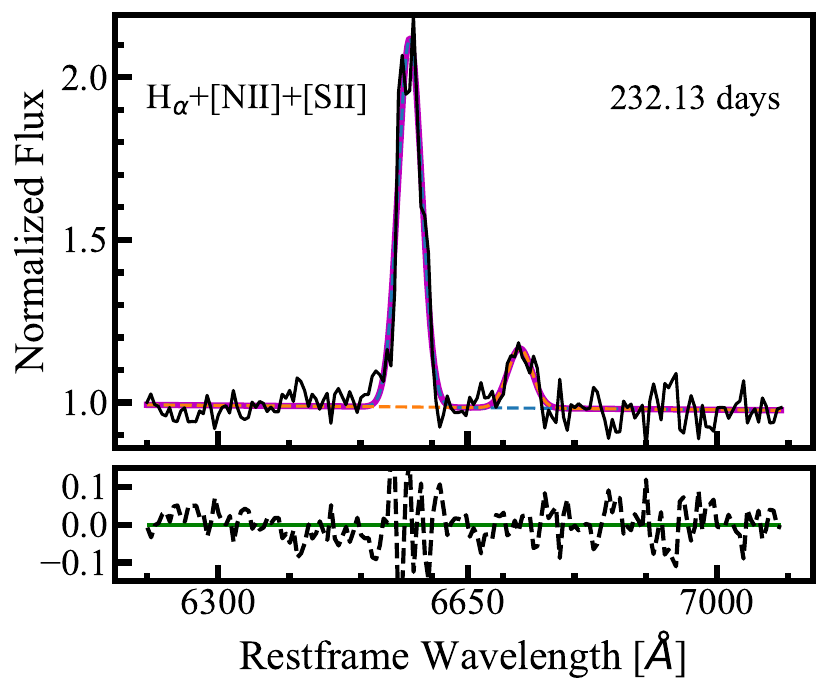}
\caption{Results from the AT\,2022wtn emission lines fit in the \ion{He}{II}+H$\upbeta$ and \ion{He}{I}+H$\upalpha$ regions (left and right panels, respectively) for 6 different phases: 23.66, 49.58, 59.91, 79.78, 105.96 and 232.13 days from the transient discovery. Starting from 105.96 days no \ion{He}{I} broad component is found. The Gaussian components are shown with dashed colored lines while the total model is represented with the magenta solid line. At the bottom of each panel, the residuals with respect the fitting model are shown. The location of the \ion{N}{III}$\uplambda$4640, \ion{He}{II}$\uplambda$4686 and H$\upbeta$ is marked with the colored dotted vertical lines (orange, cyan and black, respectively).}
\label{fig:line_fit}
\end{figure}

Another indication for the presence of an outflow is represented by the additional very broad feature detected in the \ion{He}{II}+H$\upbeta$ region in the spectrum taken at 79.78 (see the central left panels of Figure \ref{fig:line_fit}). Specifically, it is detected at a central wavelength $\uplambda_c$=(4731.68$\pm$38.88)\AA\/ (marked with a grey/green cross in the central left panel of Figure \ref{fig:fwhm_evolution}) and with a FWHM$\sim$2.3$\times$10$^{4}$ \kms\/(marked with a black cross in the upper-right panel of Figure \ref{fig:fwhm_evolution}). Given the complexity of this wavelength region, which at this stage is characterized by the simultaneous presence of many intermediate, broad and narrow components, it is difficult to give a secure interpretation for this feature. However, if we assume that it is real, it will results in very broad component 
detected with a strong red-shift (v$\sim$3$\times$10$^{3}$ \kms\/ if identified with the \ion{He}{II}$\uplambda$4686, or v$\sim$6$\times$10$^{3}$ \kms\/ if instead we ascribe this feature the the \ion{N}{III}$\uplambda$4640, grey cross and light green cross in Figure \ref{fig:fwhm_evolution}, respectively). 
Although we cannot be confident about whether this component is real, if it is real and associated with either \ion{He}{II}$\uplambda$4686 or the \ion{N}{III}$\uplambda$4640 it could be interpreted as further evidence for an outflow. This interpretation is strengthened by the detection of a similar feature in the H$\upbeta$ in the same spectrum (yellow cross in Figure \ref{fig:fwhm_evolution}) and by the subsequent radio detection with the VLA (grey dashed vertical lines in Figure \ref{fig:fwhm_evolution}).

\begin{figure}
\includegraphics[width=\columnwidth]{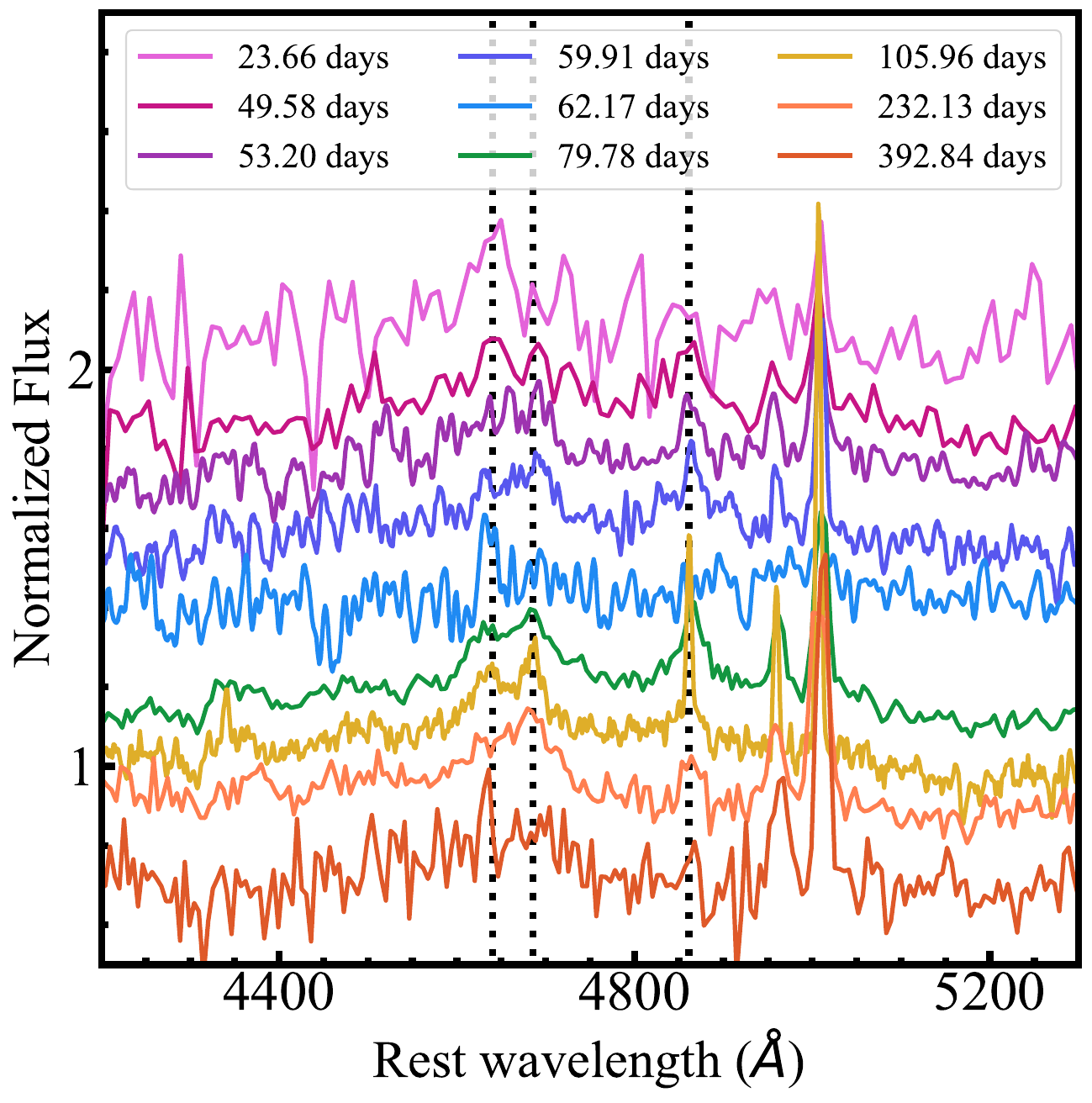}
\caption{AT\,2022wtn spectral evolution in the \ion{N}{III}+\ion{He}{II}$\lambda$4686 wavelength region. Starting from 59.91 days from the transient discovery the appearance of two separate components ascribable to the \ion{N}{III}$\lambda$4640 and \ion{He}{II}$\lambda$4686 is clearly visible. Dotted vertical lines indicate the \ion{N}{III}$\lambda$4640, \ion{He}{II}$\lambda$4686 and H$\upbeta$ position.} 
\label{fig:NIIIevolution}
\end{figure}

\section{The hosting environment}
\label{sec:host}
In this section we analyze the properties of the whole hosting environment of AT2022wtn, including the neighbor galaxy in interaction, which makes this environment particularly interesting for the case of a TDE.

\subsection{The SED fitting}
\label{subsec:sed}
In order to model the SED for each galaxy, we used the archival photometry reported in Table \ref{tab:phot_host}. Specifically, because of resolution and blending issues between the two neighbouring galaxies, the host of AT\,2022wtn SED has been fitted by using only the archival SDSS (DR16) data, while for the neighbouring galaxy, SDSS\,J232323.37+104101.7, we have been able to use also the 2 Micron All Sky Survey \cite[2MASS,][]{skrutskie06} archival photometry. 

\begin{figure*}
\centering
\includegraphics[width=1\columnwidth]{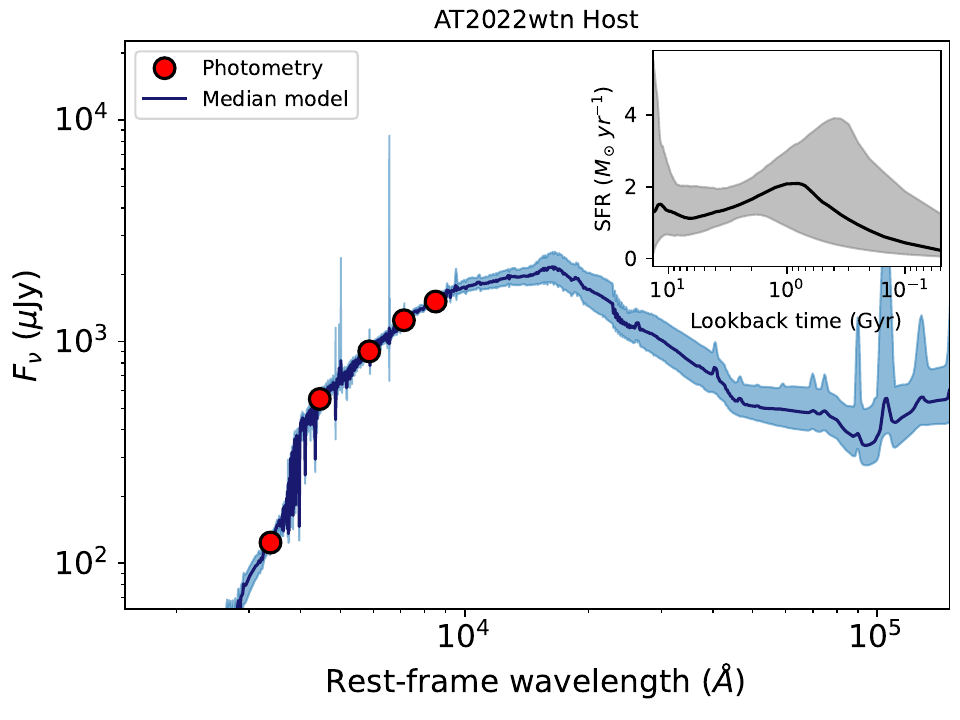}
\includegraphics[width=1\columnwidth]{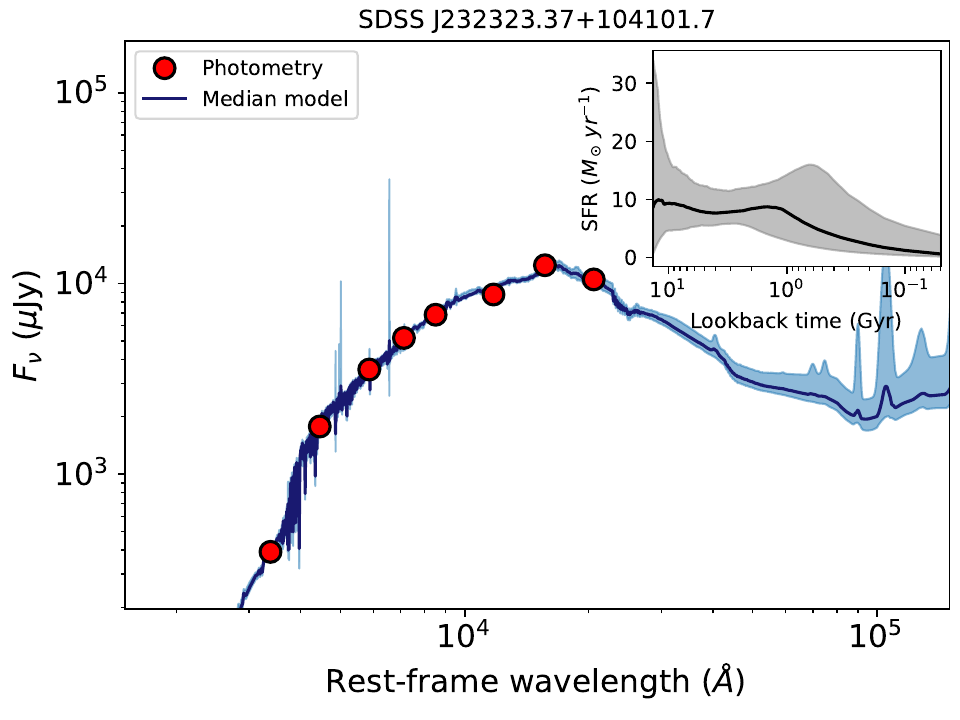}
\caption{Results from the \texttt{PROSPECTOR} modelling for both the AT\,2022wtn host and the interacting galaxy, SDSS\,J232323.37+104101.7 (left and right panels, respectively). The main figures show the archival photometry (red filled circles) and the best-fit SED model (blue) with 16th and 84th model distribution percentiles (shaded). The sub-plots, show the derived median SFH (black) and 16th and 84th percentiles (shaded).}
\label{fig:SFR_SFH}
\end{figure*}

The SED has been modeled using the stellar population synthesis models in \texttt{PROSPECTOR} \cite[][]{leja17}, which allow us to derive key physical parameters for both galaxies. In our fitting procedure, we left as free parameters the stellar mass, the metallicity, a six-component non-parametric star-formation history (SFH) and dust parameters that control the fraction and reprocessing. 
The SFH parameters include the specific star formation rate (sSFR) and the widths of five equal mass bins used to compute the SFH. The results for both galaxies, including the best-fit models, the photometry and the median star formation history profiles, are shown in Figure \ref{fig:SFR_SFH}. Specifically, for the AT\,2022wtn host, we find a stellar mass of log(M$_{\star}$/\Msun)=10.29$^{+0.12}_{-0.14}$, a metallicity of log(Z/Z$_{\odot}$)=-1.06$^{+0.78}_{-0.60}$ and a specific star formation rate in the last 50 Myr of log(sSFR)=-11.53$^{+1.21}_{-1.30}$. Comparably, for the neighbouring interacting galaxy, SDSS\,J232323.37+104101.7, we find a higher stellar mass of log(M$_{\star}$/\Msun)=11.09$^{+0.11}_{-0.12}$, a metallicity of log(Z/Z$_{\odot}$)=-0.42$^{+0.36}_{-0.57}$ and a specific star formation rate in the last 50 Myr of log(sSFR)=-11.96$^{+1.31}_{-1.00}$. In the insert panels of Figure \ref{fig:SFR_SFH} we show the weighted median SFR of each age-bin (black) and the 16th and 84th percentiles of model draws (shaded) derived from the fit versus the lookback time
since the big bang for both galaxies. 

We find that at the peak of the distribution the AT\,2022wtn host galaxy is characterized by lower values of the SFR ($\sim$2 \Msun yr$^{-1}$) with respect to the case of SDSS\,J232323.37+104101.7 ($\sim$10 \Msun yr$^{-1}$). Moreover, both galaxies show a large drop in SFR in the last $\sim$Gyr, which is consistent with what found in other TDE hosts 


\subsection{The galaxies emission lines}
\begin{figure}
\includegraphics[width=1\columnwidth]{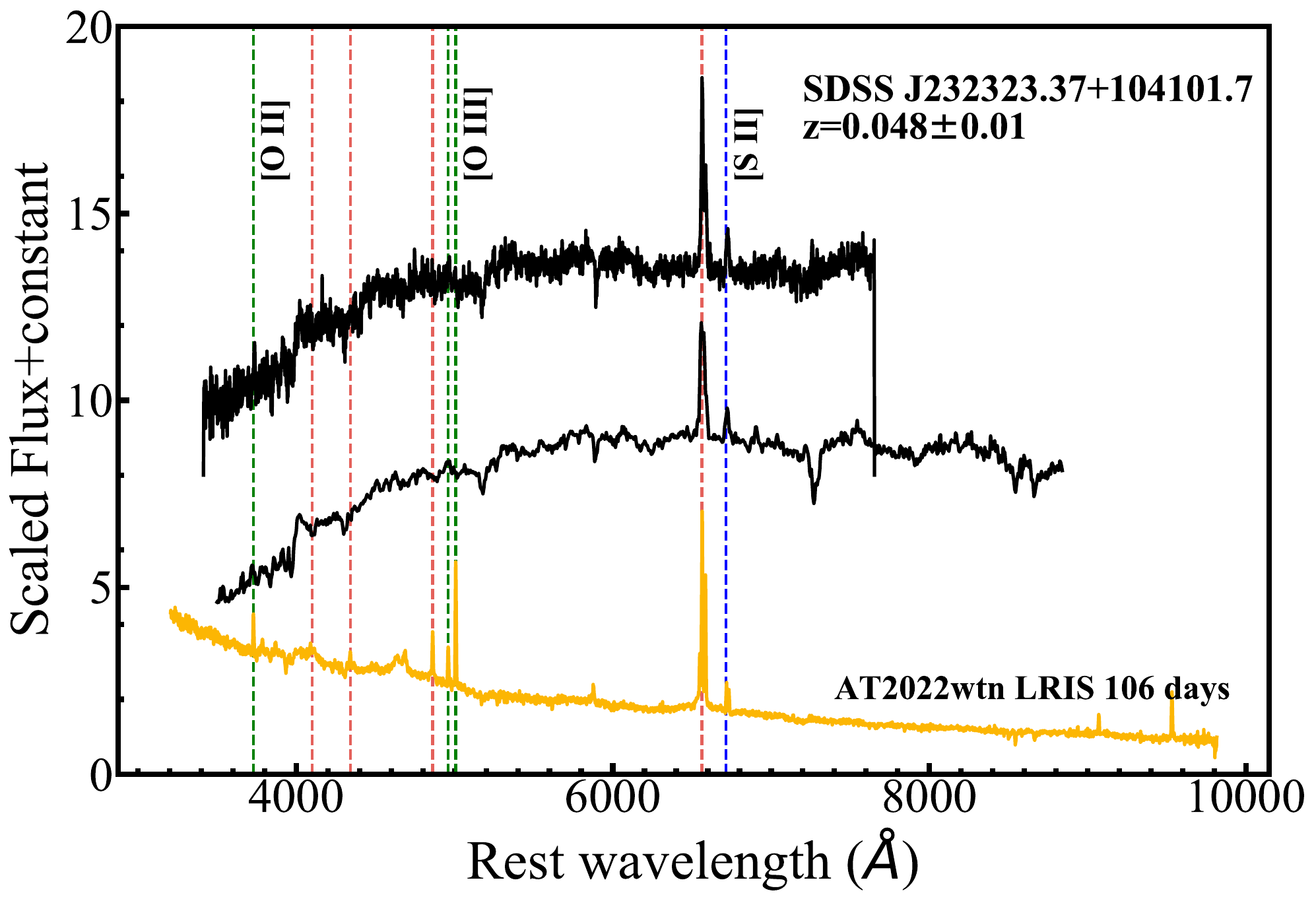}
\caption{Spectra of the neighbouring and interacting galaxy SDSS J232323.37+104101.7 (in black) in comparison with the AT\,2022wtn LRIS spectrum, taken after 105.96 days from the transient discovery (in yellow). The two galaxies are at a compatible redshift. Vertical colored dashed lines show the position of the main emission lines at this redshift: Hydrogen (red), Oxygen (green) and the [\ion{S}{II}] doublet.}
\label{fig:interacting_gal_spec}
\end{figure}

Given the peculiar hosting environment of AT\,2022wtn, we obtain also a total of two spectra of the neighbour galaxy SDSS J232323.37+104101.7 (see Table \ref{tab:spec_log}), shown in Figure \ref{fig:interacting_gal_spec} in comparison with the LRIS spectrum of AT\,2022wtn, taken at 105.96 days from the transient discovery. In both the SDSS J232323.37+104101.7 optical spectra H$\upalpha$ and [\ion{S}{II}] doublet narrow emission lines are well detected. However, the [\ion{S}{II}] doublet is not deblended, while only in the DeVeny spectrum the H$\upalpha$ is well separated from the [\ion{N}{II}]$\uplambda$6583. Thus we used the H$\upalpha$ in order to measure the redshift of this galaxy. We obtained a z=0.048$\pm$0.01, which is consistent with that of AT\,2022wtn.   

\begin{figure}
\includegraphics[width=1\columnwidth]{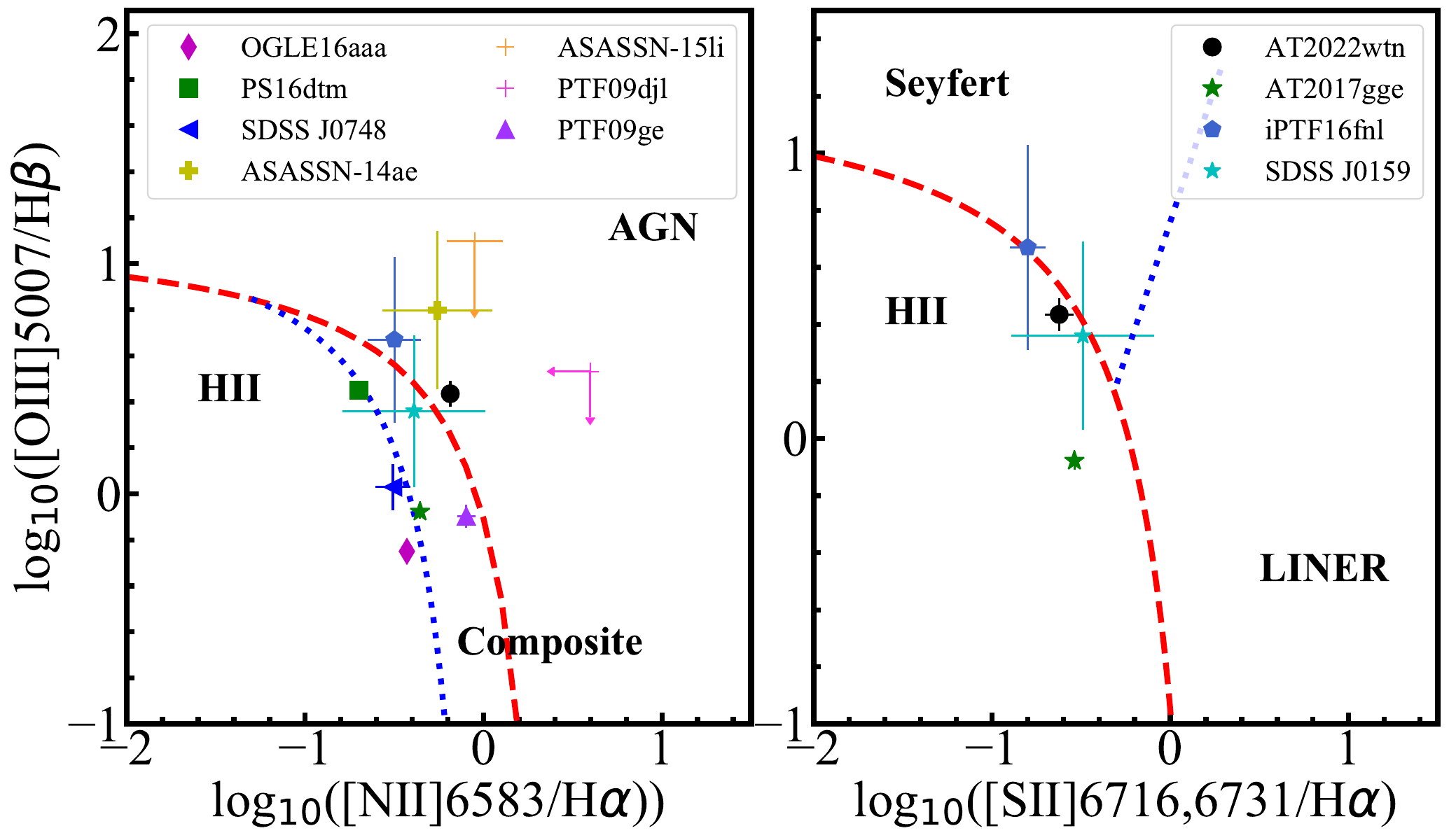}
\caption{BPT diagrams for the host of AT\,2022wtn. We used the equivalent widths of the narrow emission lines detected in the late-time Keck spectrum which we ascribe to the host environment (black filled point). The different activity regions of the diagram are separated by the following lines: red dashed line from \citep{kewley01}, blue dotted line from \citep{kauffmann03} in the left panel and blue dotted line from \citep{kewley06} in the right panel. For comparison we also show the position of some TDE host galaxies: : OGLE16aaa \citep{wyrzykowski17}; PS16dtm \citep{blanchard17}; SDSS J0159+0033 \citep{merloni15}; SDSS J0748; ASASSN-14ae; ASASSN-15li; PTF09djl; PTF09ge \citep{french17}; iPTF16fnl \citep{onori19}; AT\,2017gge \citep{onori22}.
}
\label{fig:BPT}
\end{figure}

Overall, the two galaxies appear of a quite different nature, with the bigger one SDSS J232323.37+104101.7 showing a face-on elliptical morphology (as deduced from visual inspection in the Legacy Survey image, shown in Figure \ref{fig:legacy}) and a quite passive spectrum with only the H$\upalpha$, [\ion{N}{II}] and [\ion{S}{II}] lines detected in emission. Instead, although still contaminated by the TDE emission in the blue part of the spectrum, the AT\,2022wtn host galaxy is characterized by the presence of a number of narrow emission lines, ascribable to the host environment. In particular, the LRIS spectrum is the one with the highest resolution we have, and, thus, it has been possible to deblend the narrow emission lines used for the BPT diagrams \cite[][]{baldwin81}. In Section \ref{sec:linefit}, we describe the multi-gaussian fitting procedure applied on the continuum-subtracted spectra of AT\,2022wtn. The same method has been used in the case of the LRIS spectrum, where both the TDE broad lines and the host galaxy narrow emission lines have been modelled separately (see Figure \ref{fig:line_fit}). From the derived EW of the relevant lines for the BPT diagnostic diagram we obtain the following ratios: log$_{10}$([\ion{N}{II}]6583/H$\upalpha$)=-0.19$\pm$0.026; log$_{10}$([\ion{O}{II}]5007/H$\upbeta$) =0.43$\pm$0.06 and log$_{10}$([\ion{S}{II}]6716,6731/H$\upalpha$)=-0.62$\pm$0.08. In Figure \ref{fig:BPT}, we show the position in the BPT diagnostic diagrams of AT\,2022wtn. For comparison we show also the results for some TDEs for which line ratios are available in literature: OGLE16aaa \cite[][]{wyrzykowski17}; PS16dtm \cite[][]{blanchard17}; SDSS J0159+0033 \cite[][]{merloni15}; SDSS J0748; ASASSN-14ae; ASASSN-15li; PTF09djl; PTF09ge \cite[][]{french17}; iPTF16fnl \cite[][]{onori19}; AT\,2017gge \cite[][]{onori22}.
The line ratios derived for AT\,2022wtn place its host galaxy slightly above the separation line between the composite and the AGN area and slightly below the separation line between the starforming and the Seyfert galaxies. Although at this phase the contribution of the TDE emission features is non negligible, this is still a strong indication of a non-passive nature of the AT\,2022wtn host galaxy, possibly a starforming environment, which is known to be enriched in gas and dust. 
Thus, similarly to what has been observed in the TDE AT\,2017gge and AT\,2019qiz \cite[][]{onori22, short23}, a late-time development of high ionization coronal emission lines could be observed in future spectroscopy.

\subsection{Properties of the neighbouring galaxy SDSS J232323.37+104101.7}
\label{subsec:othergal}
\begin{figure}
\includegraphics[width=1\columnwidth]{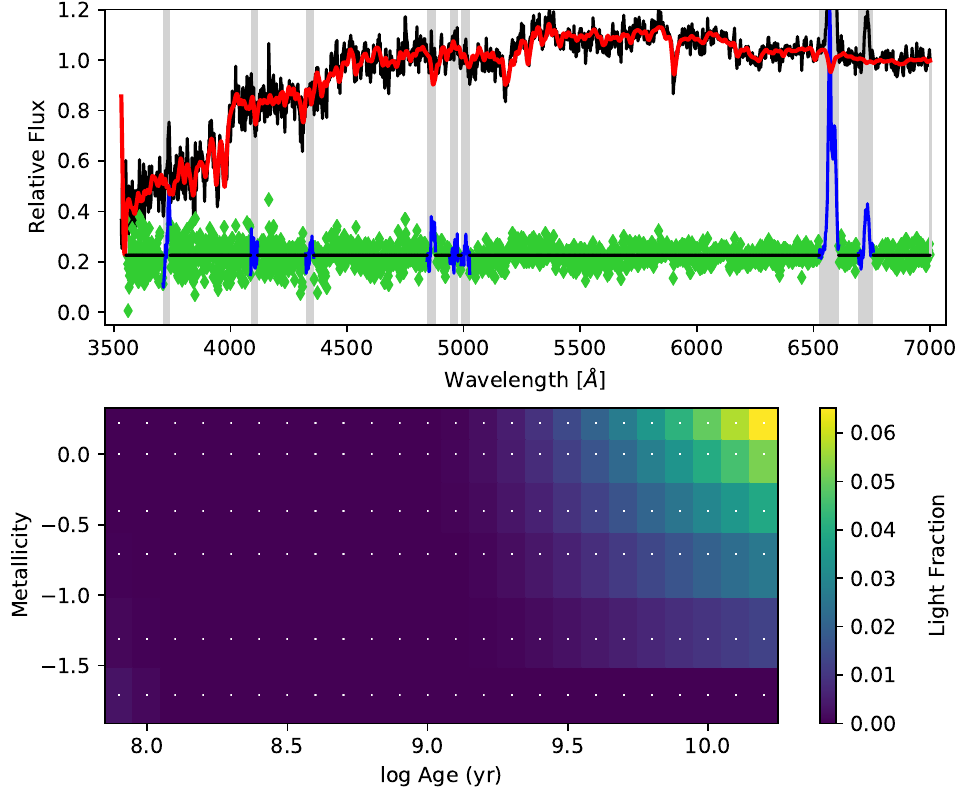}
\caption{pPXF modeling of the spectrum of neighbouring galaxy SDSS J232323.37+104101.7}
\label{fig:ppxf_othergal}
\end{figure}

Figure \ref{fig:ppxf_othergal} shows the $pPXF$\footnote{This is a Python implementation of the Penalized PiXel-Fitting method to perform full-spectrum fitting to extract the stellar and gas kinematics, as well as the stellar population of stars and galaxies (\citealt{cappellari17} and references therein).} analysis of the neighbouring galaxy's DeVeny spectrum, having approximate an instrumental resolution of R$\sim$2000 (i.e., $\sim190$ km\,s$^{-1}$). The upper panel represents the stellar and gas kinematics, showing the modeled stellar component (red curve) of the observed galaxy spectrum, corrected for redshift and galactic reddening (black curve). The galaxy emission regions have been masked (marked as shaded vertical regions) while fitting the model spectra. Green (Blue) points (curves) are residuals of the fitted (masked) regions. The model stellar component is a weighted average of the spectra taken from the Miles stellar libraries \citep{2006MNRAS.371..703S, 2011A&A...532A..95F}. The computed value of stellar velocity dispersion through $pPXF$ fit is $310\pm38$ km\,s$^{-1}$. Following the prescription of \citet{2005SSRv..116..523F}, this corresponds to a Black Hole mass of log(M$_{BH}$/\Msun)=9.15$^{+0.39}_{-0.41}$. The lower panel shows the distribution of the weights of the model components in the phase-space of stellar age and metallicity. The weighted-average value of the stellar age of the system is of log(age/year)$\sim9.85$, and the weighted-average metallicity is $\sim-0.2$ consistent with the results presented in \S\ref{subsec:sed}.

We used the luminosity of the H$\alpha$ emission line of the residual spectrum to compute the neighbouring galaxy's current star formation rate (SFR). The measured value of H$\alpha$ emission line (after deblending from [NII] $\lambda\lambda$6548,6584, and applying galactic extinction correction) is (2.66$\pm0.2$)$\times10^{-14}$ erg\,s$^{-1}$\,cm$^{-2}$. This corresponds to an H$\alpha$ luminosity of (15$\pm1$)$\times10^{40}$ erg\,s$^{-1}$. This indicates a current SFR of 1.2$\pm$0.8 \Msun\,yr$^{-1}$ (\citealt{1998ARA&A..36..189K} and references therein). 

\section{The black hole mass estimation}
In order to derive the mass of the BH involved in this TDE, we used two independent methods available in literature: modelling the multiband lightcurves with the Modular Open Source Fitter for Transients \cite[\texttt{MOSFit},][]{guillochon18} and the \texttt{TDEmass} tool \cite[][]{ryu20}. These two approaches are based on different assumptions about the main optical emission mechanism. In particular, \texttt{MOSFit} assumes that the energy generation is proportional to the instantaneous mass fallback rate (but is agnostic to the mechanism that converts this to radiation), while \texttt{TDEmass} considers the UV/optical light production as powered by shocks between intercepting stellar debris streams near the apocenters of the debris orbits. 

In the following we describe the results obtained by using each method.    

\subsection{The multicolor lightcurves fit with MOSFiT}
\begin{figure}
\includegraphics[width=\columnwidth]{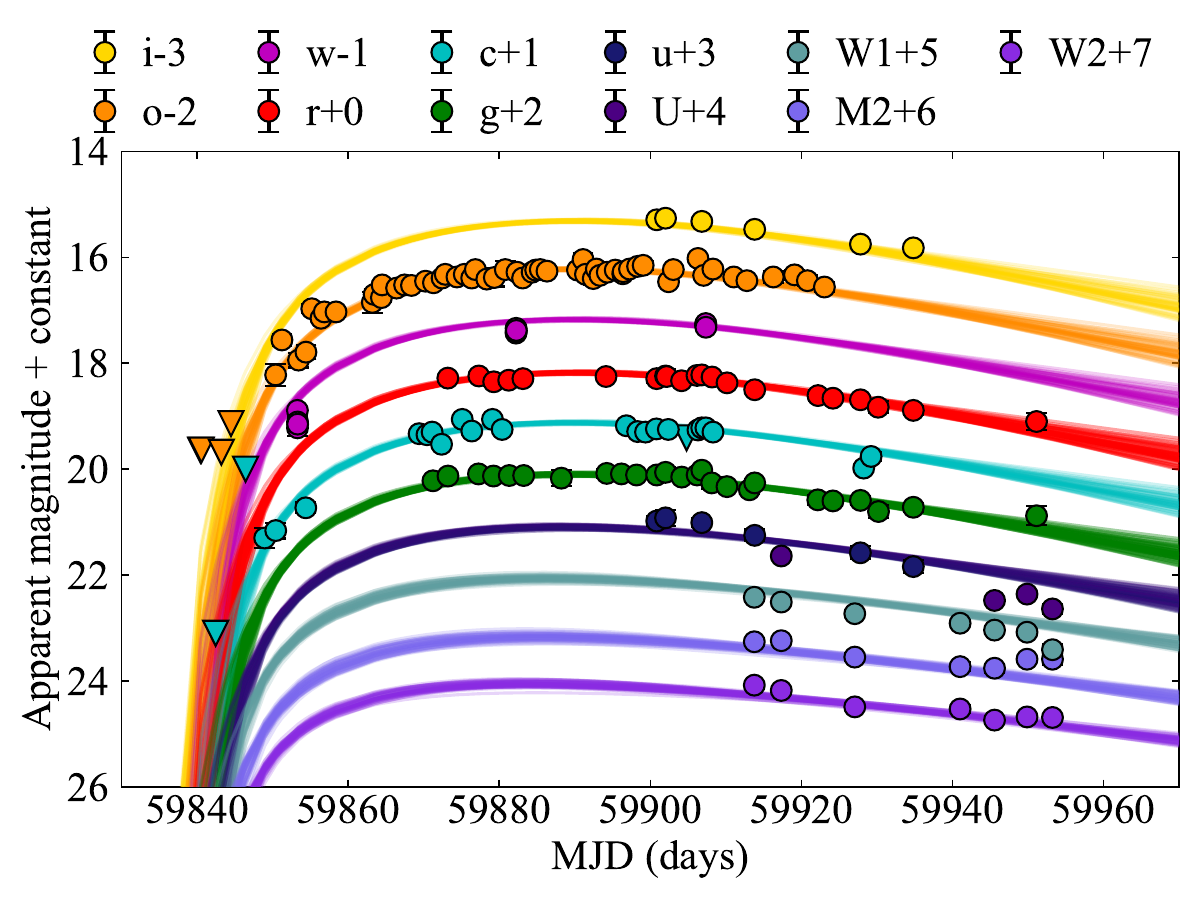}
\caption{\textsc{mosfit} fit of the AT\,2022wtn multi-colors light curves.}
\label{fig:mosfit}
\end{figure}

We derive the physical parameters of the stellar disruption by fitting the host-subtracted multi-band light curves of AT\,2022wtn by using \texttt{MOSFit} and the TDE model described in \cite{mockler19}, which assume that the rate of energy generation is proportional to the stellar debris fallback rate, whose time evolution is is taken from the simulations of \cite{guillochon13}.  
More specifically, \texttt{MOSFit} uses scaling relations and interpolations for a range of encounter parameters and masses for the BH and for the disrupted star. It generates both a bolometric light curve and multi-band light curves, which are in turn fitted to the observed data. Finally, it gives as output the combination of the highest likelihood match parameters.
The fits to the AT\,2022wtn multi-colors light curves are shown in Figure \ref{fig:mosfit}. The model represents quite accurately the photometric measurements, except for the late-time epochs (around MJD 59\,950), where the transient is too faint and it almost reached the host level. 
We find a SMBH mass of M$_{BH}$=1.2$\pm$0.2$\times$10$^{6}$ \Msun\/ and a very low mass for the disrupted star of M$_{\star}$=0.09$\pm$0.02 \Msun. Additionally, we found a scaled impact parameter b=1.32$^{+0.19}_{-0.16}$ which is consistent with a full stellar disruption \cite[for b>1, see][for details]{mockler19}.
Finally, the free parameters of the model, their priors and their posterior probability distributions are reported in Table \ref{tab:mosfit} while the two-dimensional posteriors are shown in Figure \ref{fig:mosfitCP}. We note that the posteriors on $\varepsilon$ peaks at $\sim$4\%, which is below the canonical accretion efficiency of $\sim$10\%. A possible explanation is that suggesting that a substantial fraction of the stellar mass may not be accreted, which could be connected to the radio detection.

\begin{table}
  \centering
  \begin{tabular}{cccc}
  \hline
  Parameter & Prior & Posterior & Units\\
  \hline
$ \log{(M_\bullet )}$ & $[5, 8.7]$ & $ 6.06^{+0.08}_{-0.09} $ & M$_\odot$ \\
$ M_*$ & $[0.01, 100]$ & $ 0.09^{+0.01}_{-0.02} $ & M$_\odot$ \\
$ b$ & $[0, 2]$ & $ 1.32^{+0.19}_{-0.16}$ &   \\
$ \log(\epsilon) $ & $[-4, -0.4] $ & $ -1.42 ^{+0.16}_{-0.17}$ &   \\
$ \log{(R_{\rm ph,0} )} $ & $[-4, 4] $ & $ 1.15^{+0.23}_{-0.24} $ &   \\
$ l_{\rm ph}$ & $[0, 4]$ & $ 2.39^{+0.46}_{-0.48} $ &   \\
$ \log{(T_v )} $ & $[-3, 5] $ & $ 1.23^{+0.09}_{-0.20} $ & days  \\
$ t_0 $ & $[-500, 0]$ & $  -14.56 ^{+1.96}_{-2.02}$ & days  \\
$ \log{(n_{\rm H,host})}$ & $[16, 23]$ & $ 21.02 ^{+0.04}_{-0.06}$ & cm$^{-2}$ \\
$ \log{\sigma} $ & $[-3, 2] $ & $ -0.95^{+0.03}_{-0.03}$ &   \\
  \hline
\end{tabular}
  \caption{Priors and marginalised posteriors for the \textsc{mosfit} TDE model. Priors are flat within the stated ranges, except for $M_*$, which uses a Kroupa initial mass function. The quoted results are the median of each distribution, and error bars are the 16th and 84th percentiles. These errors are purely statistical; \citet{mockler19} provide estimates of the systematic uncertainty. 
  }
  \label{tab:mosfit}
\end{table}



\subsection{The \texttt{TDEmass} fit}
\begin{figure}
\includegraphics[width=\columnwidth]{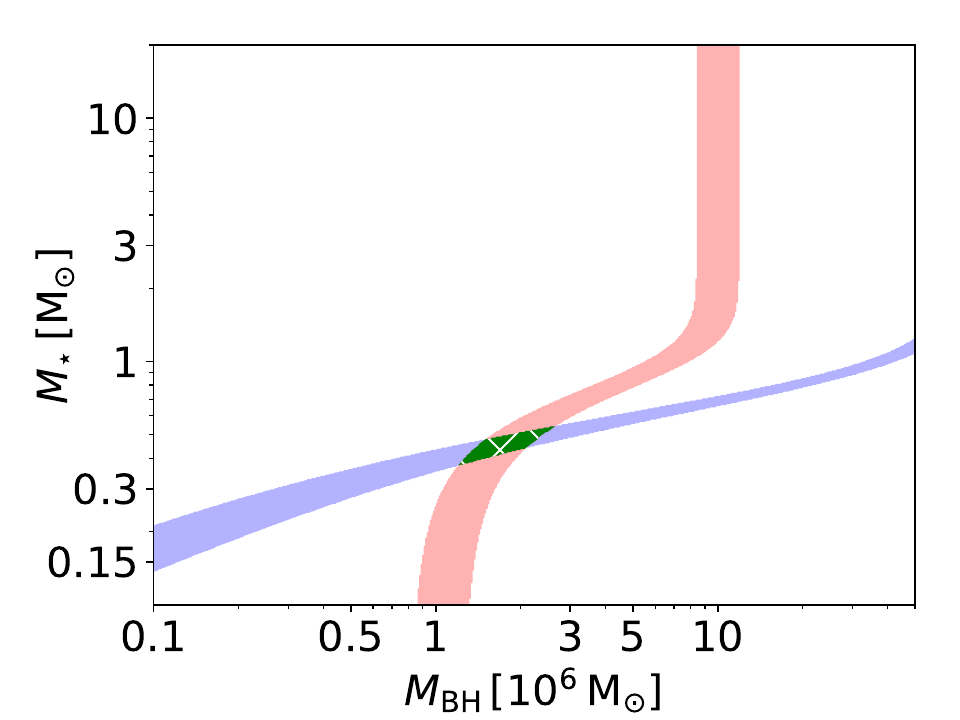}
\caption{The inferred solutions for the M$_{BH}$ and M$_{\star}$ found by using the \texttt{TDEmass} tool fot AT\,2022wtn. The blue strip indicates the solutions for the given value of the peak luminosity, while the red strip shows the solution for the given temperature input. The green region indicates the solutions found for both inputs. }
\label{fig:tdemass}
\end{figure}

\texttt{TDEmass} is a new method, recently proposed by \cite{ryu20} in order to infer the mass of the disrupted star and of the SMBH by using only two input quantities: the UV/optical luminosity and the color temperature at the peak of the flare. Differently to \texttt{MOSFit}, this tool is based on the physical model of \cite{piran15} in which the UV/optical emission originates in the outer shocks that form during the intersections of the stellar debris streams near their orbital apocenter.
For the case of AT\,2022wtn, we used as input for the peak luminosity the average of the $L_{\rm BB}$ values during the plateau phase of the TDE peak (<$L_{\rm BB,max}$>= 1.4$\pm$0.2$\times$10$^{43}$) erg/s and the corresponding averaged temperature ($T_{\rm BB}$=1.4$\pm$0.1 $\times$10$^{4}$K) for the color temperature input.
Following the tool's recommendation, we run the code keeping fixed the parameters c1 and del\_omega to the default values.
By using this independent method we obtain the following output parameters: M$_{BH}$=1.7$^{+0.9}_{-0.5}$$\times$10$^{6}$ \Msun\/ and M$_{\star}$=0.46$\pm$0.09 \Msun\/ for the mass of the BH and of the disrupted star, respectively, the characteristic mass return time of the most tightly bound debris t$_{0}$=57$^{+21}_{-15}$ days and the apocenter distance a$_{0}$=10$^{+4}_{-3} \times$10$^{14}$ cm. In Figure \ref{fig:tdemass} the inferred solution found for the M$_{BH}$ and M$_{\star}$ are shown. 
Interestingly we obtain with \texttt{MOSfit} a compatible value for the BH mass (M$_{BH}$=1.2$\pm$0.2$\times$10$^{6}$ \Msun), while, as expected from the comparison study presented in \cite{ryu20}, a smaller mass for the disrupted star (M$_{\star}$=0.09$\pm$0.02 \Msun).

\section{Discussion}
\label{sec:discussion}
\subsection{The TDE nature of AT\,2022wtn}
Our multiwavelength follow-up campaign dedicated to the transient AT\,2022wtn and covering $\sim$393 days from its discovery, enlightened a number of observable properties all compatible with a TDE nature. Specifically, its location is consistent with the nuclear region of the galaxy SDSS\,J232323.79+104107.7 (which is undergoing a merger with the more massive galaxy SDSS\,J232323.37+104101.7), 
the UV/Optical light-curve is characterized by a rising phase of $\sim$20 days, a decline consistent with the TDE models and,
as found from the photometric analysis, maximum luminosity of <L$_{\rm BB,max}$>=43.13$\pm$0.06 \unitlum\/ and a nearly constant temperature of T$_{\rm BB}\sim$1.55$\times$10$^{4}$ K. 

\begin{figure}
\includegraphics[width=\columnwidth]{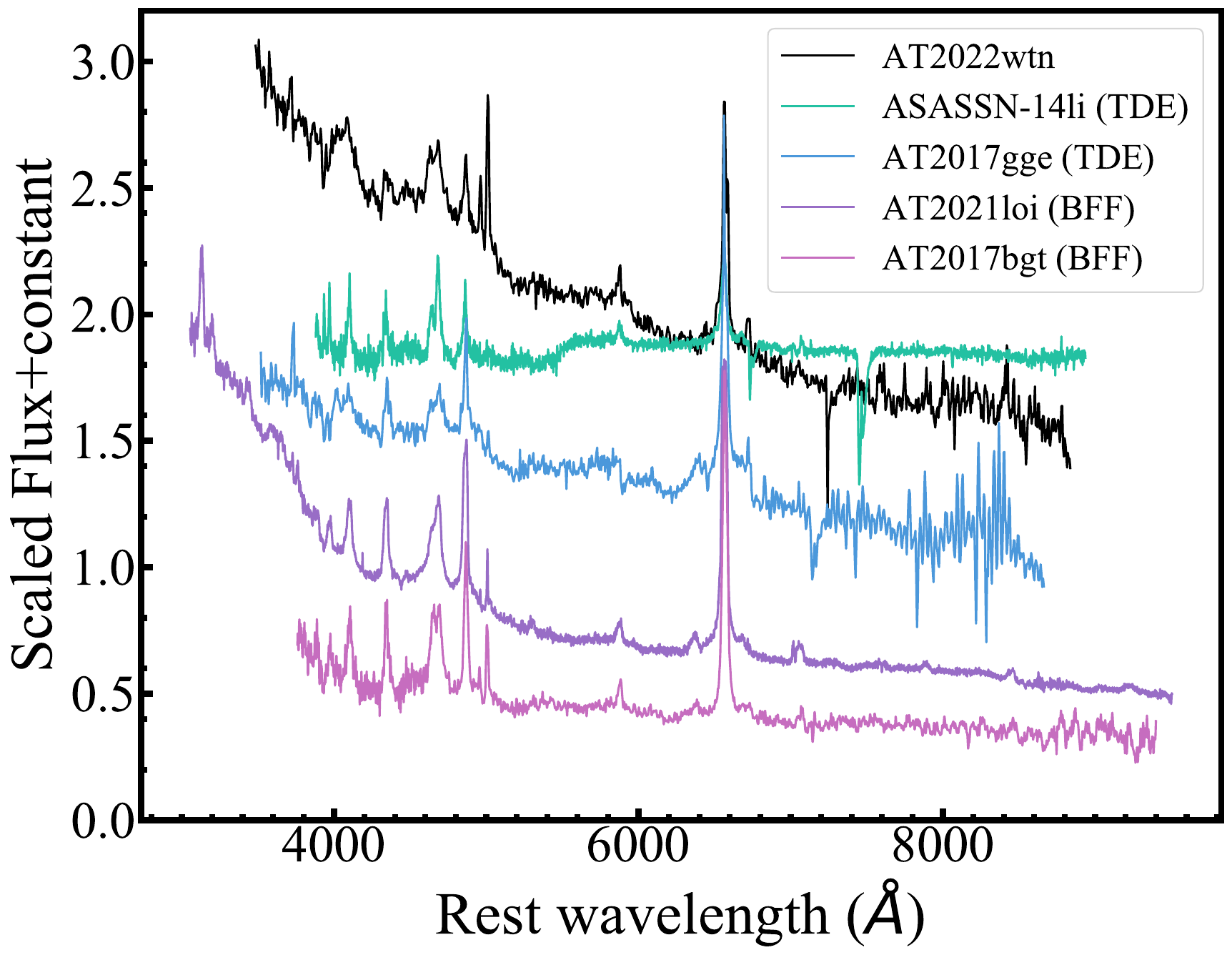}
\caption{Comparison between the AT\,2022wtn spectrum taken at 79.78 days and the optical spectra the TDEs AT\,2017gge \citep{onori22} and ASASSN-14li \citep{holoien16a} and the Bowen Fluorescence Flares (BFFs) AT\,2017bgt \citep{hosseinzadeh17} and AT\,2021loi \citep{graham2013} }
\label{fig:BFF}
\end{figure}

The X-ray non detection derived from the {\it Swift}/XRT telescope monitoring, placed AT\,2022wtn among the optically selected and X-ray faint TDE population, while the detection of broad components (FWHM$\sim$10$^{4}$ km s$^{-1}$) in the H$\upalpha$, \ion{He}{I}$\uplambda$5875 and \ion{N}{III}$\lambda$4640 in the early spectra, suggests it can be classified as a Bowen TDE-H+He subclass.
The most noticeable spectroscopic feature is the formation of the \ion{N}{II}$\uplambda$4640 + \ion{He}{II}$\uplambda$4686 double horned broad line, clearly visible starting from 59.91 days from the transient discovery. Such feature is commonly observed in the class of the Bowen Fluorescence Flare (BFF) nuclear transients, recently associated to an enhanced accretion in active SMBHs \cite[][]{makrygianni23, tadhunter17}. In Figure \ref{fig:BFF} we show a comparison between the AT\,2022wtn spectrum taken at 79.78
days and the optical spectra the two TDEs showing a similar double-horned spectral features in the \ion{He}{II}$\uplambda4686$ region, AT\,2017gge \cite[][]{onori22} and
ASASSN-14li \cite[][]{holoien16a}, and two BFFs, AT\,2017bgt \cite[][]{hosseinzadeh17, trakhtenbrot19} and AT\,2021loi \cite[][]{graham21, makrygianni23}. This spectral similarity suggests that at least some of BFFs could indeed produced by a TDE-induced rejuvenated accretion onto an active SMBH.

We estimate the physical parameters of the disruption by using two independent methods: by fitting the multi-band host subtracted light-curve with \texttt{MOSfit} and by using the \texttt{TDEmass} package. From both methods we find compatible values for the mass of the SMBH (M$_{\rm BH; \texttt{MOSfit}}$=(1.2$\pm$0.2)$\times$10$^{6}$ \Msun\/ and M$_{\rm BH; \texttt{TDEmass}}$=1.7$^{+0.9}_{-0.5}\times$10$^{6}$ \Msun) leading to an Eddington ratio range of $\uplambda_{\rm Edd}$= 0.06-0.09 (i.e. hence a sub-Eddington accretion).
In addition, the \texttt{MOSfit} analysis indicates the occurrence of a full disruption of a low-mass star (M$_{\star}$=0.09$\pm$0.02 \Msun), while the the \texttt{TDEfit} results indicate a higher mass of the disrupted star, but sill sub-solar (M$_{\star}$=0.46$\pm$0.09).
Interestingly, the low mass value of the \texttt{MOSfit} M$_{\star}$ could be explained within a scenario in which a substantial fraction of the stellar debris is not accreted because of the presence of outflows, which, in turn, causes radio emission detected at later times.

We note that finding TDEs in interacting pair of galaxies is particularly interesting given the possible implications on the rate of TDEs in merging and post-merging hosting galaxy \cite[see the results from][]{wevers24}, which we discuss in \S\ref{subsec:interactinghosts}. It appears that as a consequence of the TDEs discovery rate increasing, thanks to the development of more efficient transient dedicated surveys, we are able to enrich the TDE sample with transients in such a peculiar environments. Indeed, another confirmed TDE \cite[AT\,2023clx,][]{Charalampopoulos24} plus some TDE candidates \cite[][]{mattila18,kool20,reynolds22,payne23} have been recently discovered hosted in merging galaxies.  

Besides the observed typical TDE features, a number of interesting peculiarities have been also unveiled both from the photometric and the spectroscopic analysis which are discussed in the following  subsections, with the aim to derive clues on the TDE emission mechanism and the properties of the emitting region.   

\subsection{A TDE with a maximum luminosity plateau and a dip in the temperature}
\label{subsec:luminositydiscussion}

The first peculiar feature of AT\,2022wtn consists of a 30 days long peak in which the luminosity remains at its maximum value (<L$_{\rm BB,max}$>=43.13$\pm$0.06 \unitlum, i.e. the maximum luminosity plateau phase). During this period, we observe a corresponding decrease in the temperature, which drops to T$_{\rm BB}\sim$1.35$\times$10$^{4}$K, and remains nearly constant around this value for all the duration of the maximum luminosity plateau phase. When the luminosity starts declining, it rises again to the initial value of T$_{\rm BB}\sim$1.55$\times$10$^{4}$ K. These trends are accompanied by a fast expansion of the radius of the photosphere, characterized by a velocity of v$\sim$5000 km s$^{-1}$ and a steep rise to the maximum (R$_{\rm BB} \sim$9$\times$10$^{14}$ cm). Although steeper than the luminosity rising phase, the photosphere expansion last more days, with the radius reaching its maximum value around 10 days after the beginning of the maximum luminosity plateau. Interestingly, during this phase we detect in the spectra only the \ion{N}{III}, the H$\upalpha$ and, similar to AT\,2023clx \cite[][]{Charalampopoulos24}, the \ion{He}{I}$\lambda$5875 broad/intermediate emission lines. It is only after $\sim$10 days from the start of the luminosity decline, when the temperature is already back to its initial value and the photospheric radius shrinks, that the broad/intermediate components in the \ion{He}{II}$\lambda$4686 and in the H$\upbeta$ appear in the spectra. However, we stress that, given the very low spectral resolution of the early spectra, we may be not sensitive to the presence of a faint \ion{He}{II}$\uplambda$4686 component in blend with the stronger and broad \ion{N}{III}$\uplambda$4640.


We note that a velocity of the expanding photosphere of $\sim$5000 km s$^{-1}$ is quite fast with respect to what derived for others TDEs (specifically 2200 km s$^{-1}$ \cite[AT\,2019qiz;][]{nicholl20}, 2900 km s$^{-1}$ \cite[AT\,2020zso;][]{wevers22} and 1300 km s$^{-1}$ \cite[AT\,2020wey;][]{charalampopoulos23}) but it is similar to the case of AT\,2023clx \cite[][]{Charalampopoulos24}. Another interesting similarity between these two TDEs is that also in AT\,2023clx a 10-30 days post peak dip in temperature lasting about 30 days was observed in the photometric analysis, although there is no a corresponding maximum luminosity plateau phase. Moreover, unlike AT\,2023clx there is no NUV break detection in the AT\,2022wtn light-curves, we do not observe a double peak in the radius evolution and the rising phase of the light-curve is slower.

Following the study of \cite{wong22} on the effects of disc formation efficiency in TDEs, \cite{Charalampopoulos24} explained the photometric properties observed in AT\,2023clx with a full disruption of a low-mass star and a prompt and efficient disk formation scenario. Given that also in the case of AT\,2022wtn the photometric analysis is compatible with the full disruption of a low mass star, placing AT\,2022wtn in similar condition for the efficiency of disk formation within the \cite{wong22} analysis, we suggests that also in this case an efficient circularization process took place and led to the prompt formation of an accretion disk. The non detection of X-ray emission, but the identification of \ion{N}{III} emission lines in the early spectra, coincident with the beginning of the maximum luminosity plateau phase, are a strong indication that EUV/X-ray photons are indeed emitted from the newly formed disk, but an edge-on view of the system prevents the X-rays detection \cite[see the TDE unification model of][]{dai18}. 

The energy released by the prompt circularization and accretion is also expected to be trapped in a promptly launched quasi-spherical envelope of outflowing material (i.e. a reprocessing photosphere responsible for the UV/optical emission) \cite[][]{metzgerstone16, jiang16, lu_bonnerot20, roth20, metzger22}. Interestingly, outflows models driven primarily by circularization predict a photospheric temperature evolution similar to the dip in temperature observed in AT\,2022wtn \cite[first declining and then, after the peak of the light-curve, it rises again, see][and reference threrein]{roth20}. Moreover, if the peak in the observed luminosity is powered by the trapped radiation which adiabatically transfers energy to the gas, a longer trapping phase may explain the observed luminosity plateau in AT\,2022wtn. The contraction of the photospheric radius may mark the moment in which the mass fallback rate drops and the radiation finally can escape.

\subsection{The spectroscopic evolution and the detection of outflows}

Thanks to our promptly started spectroscopic follow-up campaign, we have been able to monitor the TDE evolution since 5 days after the transient discovery, with the first spectrum taken during the rising phase of the light-curve. It is interesting to note that broad components consistent with the TDE emission arise in the spectra only $\sim$20 days after the transient discovery, an epoch coincident with the beginning of the maximum luminosity plateau phase and the dip in the photospheric temperature. The broad features that are detected at this stage are in correspondence of the Bowen fluorescence line \ion{N}{III}$\uplambda$4640 (FWHM$\sim$8.4$\times$10$^{3}$ \kms), the  H$\upalpha$ (FWHM$\sim$14$\times$10$^{3}$ \kms) and the \ion{He}{I}$\uplambda$5875 (FWHM$\sim$6.6$\times$10$^{3}$ \kms), making AT\,2022wtn classifiable as a N-strong TDE.
It is only after the end of the maximum luminosity that we have been able to detect the \ion{He}{II}$\uplambda$4686 in the spectra, arising beside the \ion{N}{III}$\uplambda$4640 as a double-horn feature (its developments is shown in  Figures \ref{fig:line_fit} and \ref{fig:NIIIevolution}). We also note that, in contrast to what is observed in the Hydrogen and \ion{N}{III} lines, the \ion{He}{II}$\uplambda$4686 never reaches the typical width expected in the TDE broad features, it does not show velocity evolution, keeping a nearly constant width with values around FWHM$\sim$4$\times$10$^{3}$ \kms\/ during all its presence in the spectra, and it is a persistent feature, being still detected 392 days from the transient discovery (together with the \ion{N}{III} lines, see Figure \ref{fig:fwhm_evolution}). Furthermore, as shown in Figure \ref{fig:fwhm_evolution}, the detection of the \ion{He}{II}$\uplambda$4686 is soon followed by the disappearance of the \ion{He}{I}$\uplambda$5875. 

A similar spectroscopic behaviour, coupled with a similar temperature evolution is peculiar, but not unique. Indeed it has recently been observed also in the TDEs AT\,2018hyz and AT\,2023clx  \cite[][]{gomez20,short20, Charalampopoulos24}, where, in both cases a temperature dip has been observed to be accompanied with the early presence of \ion{He}{I} lines and a late-time emergence of the \ion{He}{II} in correspondence of the temperature rising phase. However, it is worth noting that, unlike AT\,2022wtn, in both these two TDEs the pseudo-bolometric luminosity does not show a maximum luminosity plateau phase.  

In the case of AT\,2023clx, \citet{Charalampopoulos24} ascribe the early presence of the \ion{He}{I} to the temperature drop as it allows the \ion{He}{II} to recombine into \ion{He}{I}. As the temperature rises again \ion{He}{I} gets ionized and the \ion{He}{II} arises again. This explanation could work also for AT\,2022wtn, although none of these features have been detected in the first spectrum. However, we note that we have no spectroscopic observations between 5 days and 20 days, the spectrum taken at 5 days is at low S/N and the low resolution of the first three spectra could prevent us from de-blending a faint \ion{He}{II}$\uplambda$4686 component from the \ion{N}{III}$\uplambda$4640 broad line. Thus, it is possible, but not secure, that we are witnessing a late-time developing of the \ion{He}{II}$\uplambda$4686. However, independently from our capability to detect a possible faint \ion{He}{II} in the early spectra, the recombination of the \ion{He}{I} into \ion{He}{II} after the temperature dip can still explain the \ion{He}{II} increase in intensity at this stage as, after these epochs, we are able to de-blend its contribution from the \ion{N}{III} even in the late-time low-resolution spectra).
 
In any case, as discussed in the \S\ref{subsec:luminositydiscussion}, the AT\,2022wtn photometric properties are consistent with an efficient circularization process, a prompt accretion disk formation and the launch of a quasi-spherical reprocessing envelope of fast expanding outflowing material. The non-detection of X-ray emission from the disk, coupled with the presence of Bowen lines in the early spectra, further support this scenario (with an edge-on viewing angle for this system), given that ionization of a reprocessing atmosphere by X-ray/EUV photons is needed in order to trigger the Bowen fluorescence mechanism. A variation in density and temperature in this expanding reprocessing photosphere could be responsible for the late-time detection of the \ion{He}{II}. 

Interestingly, \cite[][]{gomez20, short20} explain the properties of \ion{He}{II} in the AT\,2018hyz spectra either by the presence of an outflowing reprocessing material or due to a late-time shocks in debris collisions, with the \ion{He}{II} produced in a region placed further out with respect to the Hydrogen lines, given that it shows a lower width, different line profile evolution with respect to the other emission lines and a late-time appearance. 
As discussed before, we observe similar \ion{He}{II} properties also in the case of AT\,2022wtn and, in addition, we have indications for the presence of a fast-expanding outflow. Indeed, as outlined in \S\ref{sec:outflows}, we detect velocity offsets in the broad/intermediate components of Hydrogen, \ion{He}{I} and \ion{N}{III} emission lines and two additional very broad features with strong red-shift in the \ion{He}{II}+H$\upbeta$ region of the spectrum taken at 79.78 days. Given that these broad features are characterized by very complicated line profiles, we tentatively ascribe these components either to \ion{N}{III} or \ion{He}{II} and to H$\upbeta$, but we cannot be secure. However, their strong shift in wavelength coupled with the detection of a radio emission of transient nature consistent with the position of AT\,2022wtn in the following days \cite[][]{christy23}, strongly support an outflows origin for these features.    




\subsection{An interacting host: implications for the TDE rate}
\label{subsec:interactinghosts}

Since two TDEs -- AT\,2022wtn and AT\,2023clx -- out of the few tens of optically selected events have now been discovered in interacting galaxies, in this section we will discuss and estimate how much more frequently TDE host galaxies undergo mergers compared to the general population of similar galaxies. We note that in the IR also a number of events have been discovered in interacting pairs of galaxies and galaxy merges suggesting the presence of a dust-obscured population of TDEs therein \cite[][]{mattila18,kool20,reynolds22}. However, in our discussion below we will focus only on optical TDEs. 

A simplified method to derive such an estimation is quite straightforward: assuming a sample size of approximately the number of known TDEs ($\sim$100) and assuming that only one TDE is in a galaxy with a double nuclei and interaction tails (i.e. AT\,2022wtn), we can derive a frequency of such events of $\sim$1$\%$. Using visual classification of a general population of galaxies, \cite[][]{bridge10} found that the fraction of galaxies with double nuclei and tails is $\sim$0.1\% at low redshift (z $<$ 0.2). This simple metric implies TDEs are 10 times more likely to be in this class of merger than the general galaxy population (i.e. a TDE boost factor of 10). Of course, finding a second TDE host in this stage, would increase the boost factor to 20, and shows that there is significant sampling uncertainty in this method due to the small number of known TDEs.

Alternatively, if we assume that AT\,2022wtn is the only TDE in an interacting system, we can compare the 1$\%$ frequency of TDEs in interacting systems to the observed merger fractions of galaxies. Typically, the observed merger fraction is measured by assuming that close pairs galaxies are bound, and will merge on some to be defined timescale. In this framework, in order to derive an estimation of the boost TDE factor we would like to get a merger fraction per Gyr from the literature for the case of our mass and mass ratio galaxy and then compare that to the implied merger fraction from AT\,2022wtn. Specifically for the case of AT\,2022wtn, we have obtained the mass of both the two galaxies involved in the interaction from the SED fitting (see \S\ref{subsec:sed}), which result to be log(M$_{\star}$/\Msun)=11.09$^{+0.11}_{-0.12}$ for the primary galaxy (SDSS\,J232323.37+104101.7) and log(M$_{\star}$/\Msun)=10.29$^{+0.12}_{-0.14}$ for the AT\,2022wtn host. Thus, the mass ratio in this system is $\mu$$\approx$1/10, making this a minor merger. According to \cite{mundy17}, for system similar to the one hosting AT\,2022wtn (minor merger and separations in the range 5  kpc $<$ r $<$ 30 kpc ) the minor merger pair fraction at z = 0 is 0.02. To turn this into a merger rate, we need to assume a timescale for this merger to occur. For pairs with this separation and mass ratio, \cite{conselice22} finds that the pair lifetime at z = 0 is 1.93 Gyrs, which would give a merger rate per Gyr of 0.02/1.93 Gyr = 0.010 Gyr$^{-1}$. This is consistent with direct numerical simulations which suggest for the AT\,2022wtn case a merger rate per Gyr of 0.007 \cite[for a 1/4 - 1/10 merger of a 10$^{10}$- 10$^{11}$ \Msun primary galaxy,][]{oleary21}.

Now, we can estimate the merger rate in the population of TDEs by assuming that only AT 2022wtn has a close companion. This is clearly a lower limit, and so will lead to a lower limit on the boost factor. AT\,2022wtn is at redshift 0.049, which means the angular separation of 8.7 arcseconds between the centers of the two galaxies is equivalent to 8.4 kpc.  From the literature on numerical simulations, we can estimate the timescale until AT\,2022wtn completes the merger. Generally, the timescales on which the kind of morphological features are seen are roughly $\sim$0.2 - 0.4 Gyrs \cite[][]{lotz10,oleary21}. Notice that this is different from the timescale of the previous paragraph because the tidal tails of the interacting galaxy live for a much shorter time.

We can determine how many galaxies should be merging in the TDE population if they are part of the general population, namely: 
\begin{equation}
N_m = N_g * \Gamma_{\rm{merg}} * \tau
\end{equation}
where $N_m$ is the number of merger pairs, $N_g$ is the number of galaxies in the sample, $\Gamma_{\rm{merg}}$ is the merger rate and $\tau$ is the merger lifetime. If we assume a sample of TDEs ($N_g$) of 100, our computed merger rate $\Gamma_{\rm{merg}}$ = 0.010 Gyr$^{-1}$, and an average lifetime $\tau\approx0.3$ Gyrs, then we would expect to see $100\times0.01$ Gyr$^{-1} \times0.3$ Gyrs = 0.3 galaxies in this merger stage. This is lower (0.21 galaxies) if we use the numerical value of 0.007 Gyr$^{-1}$. Given that at least one galaxy is observed in this phase, this implies the minimum boost factor from this method is 3 - 5.  This is a conservative lower limit to the boost fraction, because it implies that there are no close galaxy pairs (within 30 kpc) among any of the other TDE hosts. Indeed, considering that a second TDE has been observed in a close galaxy pair \cite[AT2023clx,][]{Charalampopoulos24}, then the boost factor doubles to 6 - 10, consistent with our other method. Again, this is a lower limit to the boost factor, and highlights just how unlikely it is to find TDEs in interacting galaxies if they were not overrepresented.


We have clearly shown that there is an over-representation of interacting galaxies hosting TDEs. Given that TDEs are also known to be over-represented in post-starburst galaxies (E+A) \citep{french16}, it may be tempting to think that the two are connected. The origin of post-starburst galaxies is still somewhat uncertain, with several possible formation channels contributing \citep{Pawlik2018}. However, the most common formation channel, and the most common at this range of stellar masses, is thought to involve a gas rich merger, which triggers an intense burst of star formation, the growth of the central black hole and  the growth of a stellar bulge component \citep{Hopkins2008a, Pawlik2019} . Either the using up of the available gas, or the feedback from the AGN, would then cause a cessation or quenching of the star formation and the galaxy would enter a post-starburst phase \citep{Hopkins2008b, Li2023}

In this scenario, mechanisms related to the burst of star formation and the formation of the bulge are expected to enhance the TDE rate, lasting into the quenched star formation phase \citep{French2018}. During the active phases of the black hole, TDEs may be difficult to observe. However, once this activity subsides, an increased rate of TDEs should be visible well into the post-starburst phase. The overabundance of TDEs in this post-starburst phase \citep{french16, lawsmith2017, graur18}, together with the recent work from \citet{wevers24} concluded that their observed increased rate of TDEs in gas-rich post-mergers galaxy very likely indicate an increased rate in mergers, strongly supports this expectation.

However, the appearance of AT\,2022wtn suggests it is in the early stages of the merger rather than in the post-starburst phase. As shown in \textsection \ref{subsec:sed}, the galaxy's star formation rate does not indicate a recent burst, and the visual appearance with strong tidal tails does not typically occur in the post-starburst phase. Therefore, we may be observing an enhanced TDE rate before the burst/merger/AGN phase. The presence of tidal tails suggests that the galaxy structure is already influenced during the first approach, which would support this observation \citep{Patton2016}.  Thus, we may be witnessing the initial enhanced rate of TDEs in interacting galaxies before they enter the AGN/ post-starburst phase, which is consistent with the prediction in \citet{wevers24}. We note that based on mid-IR observations from the Wide-field Infrared Survey Explorer (WISE) satellite \citep{reynolds22} have also suggested a strongly enhanced TDE rate in luminous and ultraluminous infrared galaxies which are often undergoing a major galaxy merger. 


\section{Conclusions:} 
\label{sec:conclusions}

\subsection{A toy-model for AT\,2022wtn}
AT\,2022wtn shows photometric and spectroscopic properties all consistent with a X-ray faint TDE belonging to the TDE-H+He Bowen subclass and it can be explained with the full disruption of a low-mass star (with mass in the range M$_{\star}\sim$0.1-0.5 \Msun) by a M$_{\rm BH}$=10$^{6}$ \Msun\/ SMBH. On the basis of a number of peculiar behaviours enlightened both from the photometric analysis and from the spectroscopic monitoring campaign, we suggest a possible toy-model for the AT\,2022wtn emission mechanism and for the properties of the emitting region. The full disruption of the low-mass star by the SMBH trigger a prompt circularization phase, leading to an efficient accretion disk formation around the SMBH and the  
launch of a quasi-spherical fast expanding outflowing reprocessing envelope. These fast outflows could have caused the non-accretion of a substantial fraction of the stellar debris. Indication for the presence of such an outflowing material emerged both in the photometric and spectroscopic analysis, where a fast expanding photosphere and complex line profiles are seen before the detection of a radio emission of transient nature consistent with the AT\,2022wtn location. 
The formation of the new accretion disk produces EUV/X-ray emission, which are not directly observed. However, the detection of Bowen fluorescence lines can be used to infer the EUV/X-ray emission already at early times, together with the presence of a reprocessing envelope and, thus, in the framework of the TDE unified model of \citet{dai18}, an edge-on view of the system. At later times (after the luminosity maximum plateau) as the photosphere contracts, the \ion{He}{II} $\uplambda$4686 is detected and it is characterized by a line profile narrower than that of the \ion{N}{III} $\uplambda$4640 and of the Hydrogen lines and without showing any particular spectral evolution. This can be explained either with a spatial separation of the regions producing these lines, with the \ion{He}{II} $\uplambda$4686 emitted further out and later on with respect the Bowen and the Hydrogen lines, or by the production of the \ion{He}{II} $\uplambda$4686 in shocks from returning stellar debris streams that are revealed only at late times, when the obscuring photosphere recedes.

\subsection{A TDE in an interacting pair of galaxies}
Given the peculiar hosting environment, we have carefully investigated the properties of both the interacting galaxies through SED fitting and spectroscopic analysis. Overall, the two galaxies appear of a quite different nature, with the bigger one (SDSS\,J232323.37+104101.7) charatcerized by a face-on elliptical morphology, a quite passive spectrum (only few lines in emission have been detected: H$\upalpha$, [\ion{N}{II}] and [\ion{S}{II}]) and a higher stellar mass. Instead, the AT\,2022wtn host galaxy is characterized by the presence of a number of narrow emission lines which indicate a star-forming nature and by a lower SFR with respect to the neighbour galaxy at the peak of the distribution. However, we also find in both galaxies a large drop in the SFR in the last $\sim$Gyr which is similar to what already observed in other TDE hosts. The difference found in the stellar mass of the two galaxies results in a mass ratio of 1:10, which is consistent with a minor merger.  

Although based only on two cases (i.e. the only optical TDEs hosted in an interacting pair of galaxies system discovered so far), we have shown that there is indication for an over-representation of merging galaxies hosting TDEs, with an estimated increase of the TDE rate in such systems by a factor of 10. This results is also supported by the recently observed  increased rate of TDEs in gas-rich
post-mergers galaxy which in turn very likely indicate an increased rate in mergers \cite[][]{wevers24}. Specifically for the case of AT\,2022wtn, the galaxy’s star formation rate and its visual appearance with strong tidal tails, suggest that it is in the early
stages of the merger rather than in the post-starburst phase and that the galaxy structure has been already
influenced during the first approach. Thus, we may be indeed witnessing the initial enhanced rate of TDEs before these two galaxies enter in the 
AGN/post-starburst phase. 

This fascinating scenario needs a larger TDE host galaxies sample to be further investigated and the activity of extremely efficient transient survey such as the upcoming  Rubin Observatory Legacy Survey of Space and Time (Rubin/LSST), with the predicted increase of the TDE detection rate from 10 to $\sim$1000 events per year \cite[][]{bricman20}, will surely help in resolving the question of the TDE over-representation in merging galaxies in a framework of dedicated statistical studies.   

\section*{Acknowledgements}
This work is dedicated to my daughter Delia, who was with(-in) me all the time, during the data analysis, interpretation and writing phases that led to the birth of this manuscript.

We thanks the anonymous referee for the useful comments that contribute to the improvement of the manuscript. We thank Daniel Perley for his contribution to the spectroscopic dataset with the LT data.
This work makes use of observations from the Las Cumbres Observatory global telescope network. 
The Liverpool Telescope is operated on the island of La Palma by Liverpool John Moores University in the Spanish Observatorio del Roque de los Muchachos of the Instituto de Astrofisica de Canarias with financial support from the UK Science and Technology Facilities Council.
These results made use of the Lowell Discovery Telescope (LDT) at Lowell Observatory.  Lowell is a private, non-profit institution dedicated to astrophysical research and public appreciation of astronomy and operates the LDT in partnership with Boston University, the University of Maryland, the University of Toledo, Northern Arizona University and Yale University. Based on observations collected at the European Organisation for Astronomical Research in the Southern Hemisphere, Chile, as part of ePESSTO+ (the advanced Public ESO Spectroscopic Survey for Transient Objects Survey). ePESSTO+ observations were obtained under ESO programs ID 111.24PR and 112.25JQ.
Some of the data presented herein were obtained at the W. M. Keck Observatory, which is operated as a scientific partnership among the California Institute of Technology, the University of California and the National Aeronautics and Space Administration. We wish to recognize and acknowledge the cultural role and reverence that the summit of Maunakea has always had within the indigenous Hawaiian community. 
The SED Machine is based upon work supported by the National Science Foundation under Grant No. 1106171.
ATLAS is primarily funded to search for near earth asteroids through NASA grants NN12AR55G, 80NSSC18K0284, and 80NSSC18K1575. Science products are made possible by grants Kepler/K2 J1944/80NSSC19K0112,HST GO-15889, and STFC grants ST/T000198/1 and ST/S006109/1 and contributions form University of Hawaii Institute for Astronomy, the Queen’s University Belfast, the Space Telescope Science Institute, the South African Astronomical Observatory, and The Millennium Institute of Astrophysics (MAS), Chile. 
Pan-STARRS telescopes are supported by the National Aeronautics and Space Administration under Grants NNX12AR65G NNX14AM74G, from the Near-Earth Object Observations Program. Data are processed at Queen's University Belfast enabled through the STFC grants ST/P000312/1 and ST/T000198/1. This work was funded by ANID, Millennium Science Initiative, ICN12$\_$009. The material is based upon work supported by NASA under award number 80GSFC21M0002. FO acknowledges support from MIUR, PRIN 2020 (grant 2020KB33TP) ``Multimessenger astronomy in the Einstein Telescope Era (METE)'' and from INAF-MINIGRANT (2023): "SeaTiDE - Searching for Tidal Disruption Events with ZTF: the Tidal Disruption Event population in the era of wide field surveys". MB, EC and TP acknowledge the financial support from the Slovenian Research Agency (grants I0-0033, P1-0031, J1-8136, J1-2460 and Z1-1853) and the Young Researchers program. TWC acknowledges the Yushan Fellow Program by the Ministry of Education, Taiwan for the financial support (MOE-111-YSFMS-0008-001-P1). PC was supported by the Science \& Technology Facilities Council [grants ST/S000550/1 and ST/W001225/1]. CPG acknowledges financial support from the Secretary of Universities and Research (Government of Catalonia) and by the Horizon 2020 Research and Innovation Programme of the European Union under the Marie Sk\l{}odowska-Curie and the Beatriu de Pin\'os 2021 BP 00168 programme, from the Spanish Ministerio de Ciencia e Innovaci\'on (MCIN) and the Agencia Estatal de Investigaci\'on (AEI) 10.13039/501100011033 under the PID2020-115253GA-I00 HOSTFLOWS project, and the program Unidad de Excelencia Mar\'ia de Maeztu CEX2020-001058-M. T.E.M.B. acknowledges financial support from the Spanish Ministerio de Ciencia e Innovaci\'on (MCIN), the Agencia Estatal de Investigaci\'on (AEI) 10.13039/501100011033, and the European Union Next Generation EU/PRTR funds under the 2021 Juan de la Cierva program FJC2021-047124-I and the PID2020-115253GA-I00 HOSTFLOWS project, from Centro Superior de Investigaciones Cient\'ificas (CSIC) under the PIE project 20215AT016, and the program Unidad de Excelencia Mar\'ia de Maeztu CEX2020-001058-M. T.R. acknowledges support from the Research Council of Finland project 350458 and the Cosmic Dawn Center (DAWN) which is funded by the Danish National Research Foundation under grant DNRF140. S.M. acknowledges support from the Research Council of Finland project 350458. RS acknowledges support from grants by the National Science Foundation  (AST 2206730) and the David and Lucille Packard Foundation (PI Kasliwal).
MN is supported by the European Research Council (ERC) under the European Union’s Horizon 2020 research and innovation programme (grant agreement No.~948381) and by UK Space Agency Grant No.~ST/Y000692/1.
I.A. acknowledges support from the European Research Council (ERC) under the European Union’s Horizon 2020 research and innovation program (grant agreement number 852097), from the Israel Science Foundation (grant number 2752/19), from the United States - Israel Binational Science Foundation (BSF; grant number 2018166), and from the Pazy foundation (grant number 216312). LM acknowledges support through a UK Research and Innovation Future Leaders Fellowship (grant number MR/T044136/1). S.G.G. acknowledges support from the ESO Scientific Visitor Programme.

\section*{Data Availability}
The photometric data from I:IO/LT, LCOGT and UVOT underlying this article are available in the article itself. The PS1, ATLAS and ZTF are publicity available. The NTT spectra are publicity available through the PESSTO SSDR4 ESO Phase 4 Data Release (see the \href{http://archive.eso.org/wdb/wdb/adp/phase3_spectral/form?phase3_collection=PESSTO&release_tag=1}{ESO archive search and retrieve interface}) and on the The Weizmann Interactive Supernova Data Repository (WISeREP, https://www.wiserep.org)  The remain processed spectroscopic data underlying this article are made available as online supplementary data. All the dataset will be also shared on request to the corresponding author.



\bibliographystyle{mnras}
\bibliography{mybib_TDE} 




\appendix
\section{Optical spectroscopy}
\label{sec:Aspec}

\begin{figure*}
\includegraphics[width=2\columnwidth]{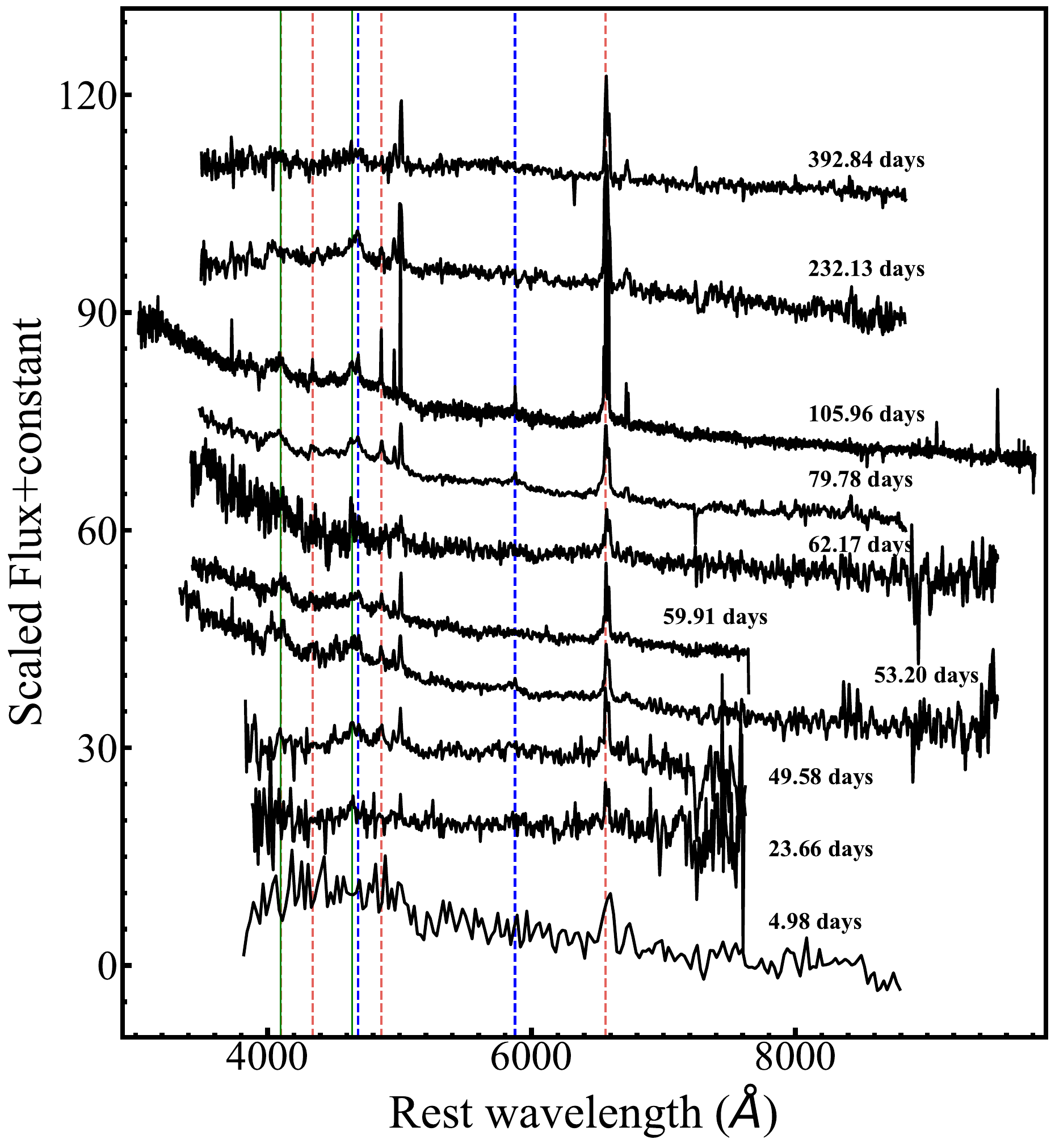}
\caption{Sequence of AT\,2022wtn rest-frame optical spectra corrected for galactic extinction. Days from the transient discovery are shown. The position of the main TDE emission lines are indicated with coloured vertical lines: Hydrogen with red dashed lines, Helium with blue dashed lines, Bowen lines with green solid lines.}
\label{fig:spec}
\end{figure*}

Here we show the spectral sequence of AT\,2022wtn starting $\sim$5 days from the transient discovery as well as the two spectra of the neighbouring galaxy SDSS
J232323.37+104101.7, in merging with the AT\,2022wtn hosting galaxy (see Figure \ref{fig:spec}). In Table \ref{tab:line_fit} we report the spectral fitting results for the \ion{N}{III}$\uplambda\uplambda$4100,4640; \ion{He}{II}$\uplambda$4686; H$\upbeta$; \ion{He}{I}$\uplambda$5875 and H$\upalpha$ emission lines.  Data observation and reduction are described in Section \ref{sec:observations}.

\begin{table*}
\fontsize{7.4}{8.4}\selectfont 
\centering 
\begin{minipage}{150mm}
\caption{Results from the emission line fit}
\begin{center}
\begin{tabular}{@{}llcccccc}
\hline
Days & Parameters &\ion{N}{III}$\lambda$4100 & \ion{N}{III}$\lambda$4510 & \multicolumn{2}{c}{\ion{N}{III}$\lambda$4640} & \multicolumn{2}{c}{\ion{He}{II}$\lambda$4686}\\ 
    &  &   &  & N & B & N & B \\
\hline
23.66 &$\uplambda_{c}$ [\AA]& $\cdots$ & $\cdots$ & $\cdots$ &4641.77 $\pm$7.27& $\cdots$& $\cdots$\\
   &FWHM [\kms] &  $\cdots$ & $\cdots$ & $\cdots$ &8404$\pm$1213 & $\cdots$& $\cdots$\\
    &EW [\AA]    & $\cdots$ & $\cdots$ &  $\cdots$ &27.0$\pm$5.0& $\cdots$& $\cdots$\\
\hline
49.58  &$\uplambda_{c}$ [\AA]&$\cdots$& 4503.46$\pm$4.34  &  $\cdots$ & 4662.36$\pm$6.31& $\cdots$& $\cdots$\\
    &FWHM [\kms]& $\cdots$&2315$\pm$716 &  $\cdots$   & 10764$\pm$1178&  $\cdots$ &$\cdots$\\
    &EW [\AA]   & $\cdots$& 4.2$\pm$1.7  &  $\cdots$  & 32.1$\pm$4.4 & $\cdots$ &$\cdots$\\
\hline 
53.20 &$\uplambda_{c}$ [\AA]  &$\cdots$ & 4518.78$\pm$0.86 & $\cdots$ & 4658.51$\pm$1.98& $\cdots$ &$\cdots$\\
   & FWHM [\kms] & $\cdots$ & 1363$\pm$137 & $\cdots$ & 8322$\pm$329 & $\cdots$&$\cdots$\\
   & EW [\AA] & $\cdots$ & 3.1$\pm$0.4& $\cdots$& 21.4$\pm$1.1 &$\cdots$&$\cdots$\\   
 \hline  
 59.91  & $\uplambda_{c}$ [\AA]  & 4113.63$\pm$3.14 & $\cdots$& 4630.85$\pm$1.37 &4667.60 $\pm$ 5.19 & 4694.38$\pm$1.91 & $\cdots$\\ 
     &  FWHM [\kms] &6795$\pm$579&$\cdots$& 667$\pm$228 & 6093$\pm$642 &1431$\pm$349 &$\cdots$\\
     & EW [\AA] & 12.4$\pm$1.4 &$\cdots$& 0.78$\pm$0.34 &10.54$\pm$1.55  &1.93$\pm$0.60&$\cdots$\\
\hline
62.17  & $\uplambda_{c}$ [\AA]  & $\cdots$& $\cdots$& 4635.01$\pm$0.54 &$\cdots$ &4693.88$\pm$2.94&$\cdots$\\
    &  FWHM [\kms] & $\cdots$&$\cdots$&1169$\pm$88 &$\cdots$&4306$\pm$510&$\cdots$\\
    & EW [\AA] & $\cdots$& $\cdots$&11.2$\pm$1.1&$\cdots$& 15.9$\pm$2.3&$\cdots$\\
\hline
79.78 & $\uplambda_{c}$ [\AA]  & 4076.10$\pm$3.76  & $\cdots$& 4627.39$\pm$3.09& $\cdots$& 4681.69$\pm$3.14& 4731.68$\pm$38.88\\
   &  FWHM [\kms] & 9133$\pm$796 & $\cdots$& 2384$\pm$414 & $\cdots$& 3625$\pm$529 & 23139$\pm$6700\\ 
   & EW [\AA] & 13.7$\pm$1.6 & $\cdots$& 4.4$\pm$1.0 & $\cdots$&8.7$\pm$1.5& 34.0$\pm$10.2\\
\hline
105.96& $\uplambda_{c}$ [\AA]  & 4081.53$\pm$3.02&4499.47 $\pm$ 6.87 & 4634.33$\pm$1.50& 4648.60$\pm$2.32&4685.76$\pm$0.58&$\cdots$\\
   &  FWHM [\kms] & 10427$\pm$676& 11585$\pm$1128& 1632$\pm$331& 7075$\pm$442&1238$\pm$109& $\cdots$\\
   & EW [\AA] &15.0$\pm$1.2&12.90$\pm$1.56 & 1.7$\pm$0.4& 13.9$\pm$1.5&2.8$\pm$0.3& $\cdots$\\
\hline
232.13& $\uplambda_{c}$ [\AA]  & 4144.06$\pm$14.55& 4513.46$\pm$9.36&$\cdots$&4619.38$\pm$10.36& 4683.74$\pm$1.42& $\cdots$\\
   &  FWHM [\kms] & 11413$\pm$3112& 6111$\pm$1955&$\cdots$&3893$\pm$1296&4344$\pm$672& $\cdots$\\
   & EW [\AA] &10.92$\pm$2.22& 4.95$\pm$1.83& $\cdots$ &6.45$\pm$2.59&14.28$\pm$2.58 &$\cdots$\\
\hline
392.84& $\uplambda_{c}$ [\AA]  & 4051.35$\pm$5.77& $\cdots$ & $\cdots$& 4602.74$\pm$20.27& 4696.22$\pm$7.41& $\cdots$ \\
   &FWHM [\kms] & 7200$\pm$1126& $\cdots$ & $\cdots$& 16135$\pm$2676& 3746$\pm$1384& $\cdots$\\
   & EW [\AA] & 14.08$\pm$3.15& $\cdots$ & $\cdots$&27.94$\pm$6.10& 5.50$\pm$2.64& $\cdots$ \\
\hline
Days& Parameters& \multicolumn{2}{c}{H$_{\upbeta}$} & \multicolumn{2}{c}{\ion{He}{I}$\lambda$5875} & \multicolumn{2}{c}{H$_{\upalpha}$}\\
& & N & B &N &B &N & B\\
\hline
23.66 &$\uplambda_{c}$ [\AA]& $\cdots$& $\cdots$ & 5878.34$\pm$14.57& $\cdots$& 6570.15$\pm$1.59&6521.99$\pm$22.42  \\
    &FWHM [\kms] & $\cdots$& $\cdots$& 6622$\pm$1911&$\cdots$& 1522$\pm$188& 13980$\pm$3409 \\
    &EW [\AA]    & $\cdots$ &$\cdots$& 13.4$\pm$5.0&$\cdots$& 16.1$\pm$2.6& 38.2$\pm$11.6 \\
\hline
49.58 & $\uplambda_{c}$ [\AA]& 4857.25$\pm$4.31 & $\cdots$& $\cdots$& 5852.87$\pm$26.13& 6571.22$\pm$0.91& $\cdots$  \\
    &FWHM [\kms] & 4333$\pm$693& $\cdots$& $\cdots$& 15072$\pm$4157& 1783$\pm$107 & $\cdots$\\
    &EW [\AA]    &12.9±2.6&$\cdots$ & $\cdots$ & 17.9$\pm$6.1& 25.1$\pm$1.9 & $\cdots$\\
\hline
53.20 &$\uplambda_{c}$ [\AA] &4863.06$\pm$0.89& $\cdots$&5876.43$\pm$0.93&5843.65$\pm$4.31& 6566.66$\pm$0.25 & 6566.02$\pm$1.25 \\
    &FWHM [\kms] &1847$\pm$132& $\cdots$& 645$\pm$120&7586$\pm$566 & 616$\pm$32& 3242$\pm$142\\
    &EW [\AA]    &5.2$\pm$0.5& $\cdots$ &1.4$\pm$0.3&13.4$\pm$1.3 & 7.6$\pm$0.5& 19.8$\pm$1.6\\
\hline
59.91 &$\uplambda_{c}$ [\AA] & 4864.47$\pm$1.05 & 4865.72$\pm$3.62 & $\cdots$ & 5881.24$\pm$22.72 & 6565.49$\pm$0.11 & 6557.55$\pm$2.63\\
    &FWHM [\kms] & 697$\pm$172 & 6280$\pm$663& $\cdots$ & 13895$\pm$3089 & 476$\pm$16 &4420$\pm$322 \\
    &EW [\AA]    & 1.6$\pm$0.5& 12.8$\pm$1.9 & $\cdots$ & 8.8$\pm$2.5&13.3$\pm$0.5&22.0$\pm$2.6 \\
\hline
62.17 &$\uplambda_{c}$ [\AA] &4893.77$\pm$1.64& $\cdots$& $\cdots$ & $\cdots$ & 6569.53$\pm$0.53 &$\cdots$ \\
    &FWHM [\kms] &1628$\pm$242& $\cdots$&$\cdots$& $\cdots$ & 641$\pm$68 &$\cdots$\\
    &EW [\AA]    &6.3$\pm$1.2 &$\cdots$& $\cdots$ &$\cdots$ &12.7$\pm$1.8&$\cdots$\\
\hline
79.78 &$\uplambda_{c}$ [\AA] & 4865.20$\pm$1.11& 4984.70$\pm$17.35 &  5874.52$\pm$1.22& 5824.52$\pm$8.99& 6568.80$\pm$0.37 & 6554.31$\pm$4.56\\
    &FWHM [\kms] & 2078$\pm$188& 10130$\pm$3005 & 1154$\pm$156 & 16752$\pm$1379 & 1767$\pm$51& 8061$\pm$660\\
    &EW [\AA]    &6.3$\pm$0.7& 9.5$\pm$5.1&3.0$\pm$0.5& 27.5$\pm$2.8& 35.0$\pm$1.3& 29.7$\pm$3.7\\
\hline
105.96 &$\uplambda_{c}$ [\AA] & 4861.48$\pm$0.09& 4835.84$\pm$8.70& 5877.54$\pm$0.30& $\cdots$& 6564.35$\pm$0.02& 6561.75$\pm$0.79 \\
    &FWHM [\kms] &430$\pm$13& 23058$\pm$1637& 525$\pm$41& $\cdots$& 309$\pm$2& 3308$\pm$100\\
    &EW [\AA]    &3.6$\pm$0.1 & 44.8$\pm$3.4 &2.2$\pm$0.2& $\cdots$& 20.7$\pm$0.2& 19.8$\pm$1.0\\
\hline
232.13 &$\uplambda_{c}$ [\AA] &4863.74$\pm$1.42& $\cdots$& $\cdots$&$\cdots$&6568.97$\pm$0.44& $\cdots$\\
    &FWHM [\kms] &1376$\pm$213 & $\cdots$& $\cdots$&$\cdots$&1781$\pm$48 & $\cdots$\\
    &EW [\AA]    & 3.1$\pm$0.6& $\cdots$&$\cdots$&$\cdots$&47.05$\pm$1.67& $\cdots$\\
\hline
392.84& $\uplambda_{c}$ [\AA]  & 4865.55$\pm$2.30 & $\cdots$& $\cdots$& $\cdots$&6566.09$\pm$10.31& $\cdots$ \\
   & FWHM [\kms] & 738$\pm$342& $\cdots$&$\cdots$&$\cdots$& 991$\pm$37&$\cdots$\\
   & EW [\AA]    & 1.52$\pm$0.93& $\cdots$&$\cdots$&$\cdots$&35.29$\pm$1.55& $\cdots$\\
\hline
\end{tabular}
\label{tab:line_fit}
\end{center}
Notes: (1) Days from the transient's discovery; (2) Fitting parameters (3)-(6) fitting Gaussian component identification (N=narrow and B=broad).  
With $\cdots$ we indicate the cases in which the component has not been fitted (no component).\\
\end{minipage}
\end{table*}

\section{UV/Optical photometry}
\label{sec:Aphot}

\begin{figure*}
\includegraphics[width=2\columnwidth]{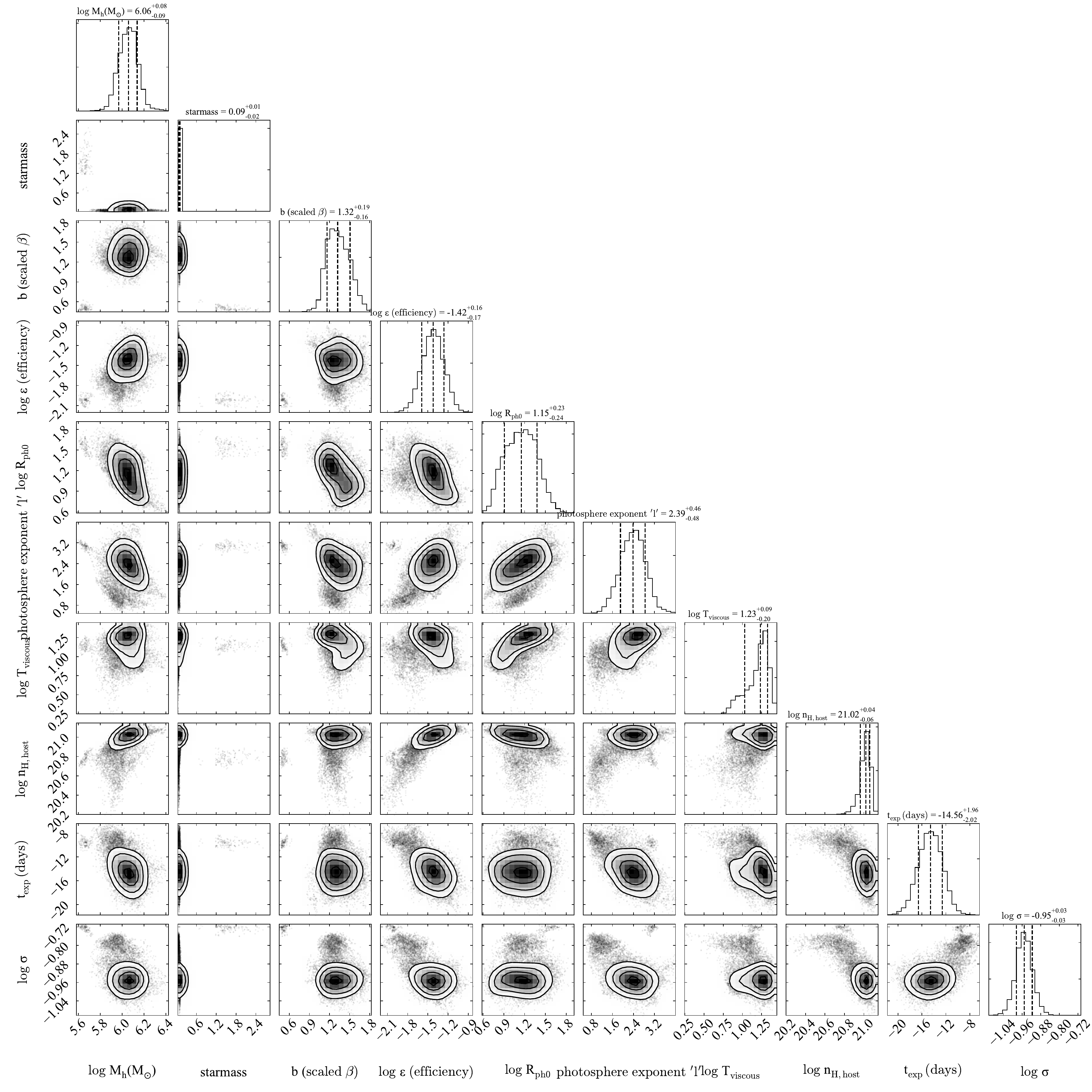}
\caption{Two-dimensional posteriors resulting from the \texttt{MOSFit} analysis on the AT\,2022wtn photometry.}
\label{fig:mosfitCP}
\end{figure*}

In the following we report the photometric measurements from the I:O/LT, LCOGT and {\it Swift}/UVOT observations (see Tables \ref{tbl:Optphot} and \ref{tbl:UVOTphot}). In Figure \ref{fig:mosfitCP} we show the two-dimensional posteriors obtained from the \texttt{MOSFit} analysis. 
\begin{table*}
\centering
\begin{minipage}{150mm}
 \caption{I:IO/LT and LCOGT photometric measurements }
 \begin{center}
 \begin{tabular}{@{}llcccc}
 \hline
 MJD       &  Phase & $u^\prime$     &$g^\prime$      &$r^\prime$     &$i^\prime$    \\
(1)        & (2)    & (3)            &(4)             &(5)            &(6)           \\
59\,900.84 & 46.58  & 17.98$\pm$0.06 & 18.11$\pm$0.03 &18.29$\pm$0.04 &18.29$\pm$0.02\\
59\,902.01 & 47.75  & 17.92$\pm$0.10 & 18.06$\pm$0.06 &18.24$\pm$0.09 &18.26$\pm$0.04\\
59\,906.09 & 51.83  & $\cdots$       & 17.99$\pm$0.11 &18.10$\pm$0.14 &18.32$\pm$0.32 \\
59\,906.10 & 51.84  & $\cdots$       & 18.03$\pm$0.10 &18.20$\pm$0.22 &18.25$\pm$0.24 \\
59\,906.81 & 52.55  & 18.01$\pm$0.04 & 18.02$\pm$0.03 &18.22$\pm$0.03 &18.32$\pm$0.03\\
59\,910.07 & 55.81  & $\cdots$       & 18.28$\pm$0.07 &18.35$\pm$0.09 &18.47$\pm$0.20\\  
59\,910.08 & 55.82  & $\cdots$       & 18.26$\pm$0.11 &18.40$\pm$0.08 &18.48$\pm$0.14\\
59\,913.81 & 59.55  & 18.25$\pm$0.07 & 18.26$\pm$0.04 &18.50$\pm$0.03 &18.47$\pm$0.02\\
59\,914.05 & 59.79  & $\cdots$       & 18.34$\pm$0.13 &18.40$\pm$0.18 &18.58$\pm$0.19\\  
59\,914.06 & 59.80  & $\cdots$       & 18.36$\pm$0.13 &18.45$\pm$0.17 &18.52$\pm$0.20\\
59\,918.08 & 63.82  & $\cdots$       & 18.35$\pm$0.18 &18.46$\pm$0.21 &18.60$\pm$0.26\\ 
59\,918.08 & 63.82  & $\cdots$       & 18.35$\pm$0.18 &18.36$\pm$0.21 &18.42$\pm$0.28\\
59\,924.05 & 69.79  & $\cdots$       & 18.39$\pm$0.09 &18.60$\pm$0.13 &18.61$\pm$0.18\\  
59\,924.06 & 69.80  & $\cdots$       & 18.41$\pm$0.09 &18.57$\pm$0.12 &18.62$\pm$0.22\\ 
59\,927.80 & 73.54  & 18.58$\pm$0.08 & 18.59$\pm$0.05 &18.69$\pm$0.04 &18.75$\pm$0.03\\
59\,928.18 & 73.92  & $\cdots$       & 18.55$\pm$0.11 &18.62$\pm$0.18 &18.89$\pm$0.32\\ 
59\,928.19 & 73.93  & $\cdots$       & 18.54$\pm$0.12 &18.65$\pm$0.19 &18.82$\pm$0.30\\ 
59\,930.43 & 76.17  & $\cdots$       & 18.71$\pm$0.15 &18.73$\pm$0.18 &19.19$\pm$0.36\\
59\,933.86 & 79.60  & $\cdots$       & 18.76$\pm$0.10 &18.88$\pm$0.14 &19.02$\pm$0.18\\
59\,934.81 & 80.55  & 18.84$\pm$0.08 & 18.72$\pm$0.05 &18.89$\pm$0.05 &18.82$\pm$0.04\\
59\,937.87 & 83.61 	& $\cdots$       & 18.82$\pm$0.12 &18.87$\pm$0.18 &18.99$\pm$0.23\\
59\,949.05 & 94.79	& $\cdots$       & 18.88$\pm$0.24 &18.98$\pm$0.26 &19.38$\pm$0.41\\
59\,953.09 & 98.83  & $\cdots$       & 18.96$\pm$0.16 &19.08$\pm$0.24 &19.07$\pm$0.34\\
59\,957.07 & 102.81	& $\cdots$       & 18.92$\pm$0.11 &19.06$\pm$0.18 &19.27$\pm$0.27\\
59\,964.08 & 109.82 & $\cdots$       & 19.08$\pm$0.12 &19.04$\pm$0.27 &19.18$\pm$0.37\\
59\,971.07 & 116.81	&  $\cdots$      & 19.11$\pm$0.26 &19.16$\pm$0.34 &$\cdots$\\
\hline
\hline
\end{tabular}
\end{center}
Notes: (1) MJD date of observations; (2) Phase (days) with respect to the discovery date MJD 59\,853.28; from (3) to (6) host-subtracted apparent magnitudes and uncertainties for the I:IO/LT and LCOGT $u^\prime$, $g^\prime$, $r^\prime$ and $i^\prime$  filters, reported in AB system. 
All the magnitudes are uncorrected for foreground extinction.
With  $\cdots$ we indicate epochs with no data available (no observations). 
\noindent
\end{minipage}
\label{tbl:Optphot}
\end{table*}

\begin{table*}
\centering
\begin{minipage}{150mm}
 \caption{List of the \textit{Swift} observations executed for the monitoring of AT\,2022wtn and the UVOT photometric measurements}
 \begin{center}
 \begin{tabular}{@{}llcccccccc}
 \hline
MJD      &  Phase & XRT exp. time & UVOT exp. time  &\textit {UVW2}  &  \textit{UVM2} & \textit {UVW1}  & \textit {U}   & \textit {B} & \textit {V}  \\
(days)   &  (days)& (s)           & (s)             & (mag)          & (mag)          & (mag)           & (mag)         & (mag)       & (mag)       \\
(1)      & (2)    & (3)           &(4)              & (5)            & (6)            & (7)             & (8)           & (9)         & (10)     \\
\hline
59\,913.76 & \phantom{1}60.5 &1631 &1543 & 16.80$\pm$0.05& 16.97$\pm$0.06& 16.94$\pm$0.05 & $\cdots$& $\cdots$ &$\cdots$\\
59\,917.33 & \phantom{1}64.0 &3935 &4021 & 16.88$\pm$0.05& 16.95$\pm$0.06& 17.00$\pm$0.06 & 16.92$\pm$0.05 & 17.49$\pm$0.04& 16.89$\pm$0.06\\
59\,927.06 & \phantom{1}73.8 &1529 &1530 & 17.10$\pm$0.06 & 17.18$\pm$0.08 & 17.13$\pm$0.07&$\cdots$& $\cdots$ &$\cdots$\\ 
59\,941.00 &\phantom{1}87.7  &1637 &1775 & 17.13$\pm$0.07 & 17.31$\pm$0.08 & 17.23$\pm$0.06&$\cdots$& $\cdots$ &$\cdots$\\ 
59\,945.56 &\phantom{1}92.3  &2694 &2641 & 17.27$\pm$0.06&17.33$\pm$0.08&17.30$\pm$0.08&17.27$\pm$0.07&17.66$\pm$0.06&16.86$\pm$0.07\\
59\,949.86 &\phantom{1}96.6  &2858 &2806 & 17.23$\pm$0.05& 17.21$\pm$0.07&17.32$\pm$0.08&17.23$\pm$0.06&17.65$\pm$0.06&16.87$\pm$0.07\\
59\,953.23 &\phantom{1}99.0  &1517 &1490 & 17.24$\pm$0.06&17.21$\pm$0.10&17.47$\pm$0.10&17.32$\pm$0.09&17.75$\pm$0.09&17.03$\pm$0.11\\ 
59\,956.61 & 102.4           &2951 &2829 & 17.22$\pm$0.06&17.34$\pm$0.09&17.30$\pm$0.09&17.21$\pm$0.08&17.55$\pm$0.07&16.90$\pm$0.09\\
59\,965.09 & 110.8           &3257 &3203 & 17.30$\pm$0.05&17.50$\pm$0.08&17.42$\pm$0.08&17.21$\pm$0.07&17.74$\pm$0.07&16.98$\pm$0.08\\   
59\,968.08 & 113.8           &1572 &1534 & 17.37$\pm$0.07&17.53$\pm$0.09&17.32$\pm$0.08& $\cdots$& $\cdots$ &$\cdots$\\ 
59\,970.85 &116.6            &3450 &3378 & 17.35$\pm$0.06&17.46$\pm$0.08&17.26$\pm$0.07&17.32$\pm$0.08&17.73$\pm$0.07&16.94$\pm$0.08\\
59\,974.03 &119.8            &4212 &4135 & 17.41$\pm$0.06& 17.46$\pm$0.08&17.44$\pm$0.08&17.17$\pm$0.07&17.64$\pm$0.07&16.97$\pm$0.09\\ 
60\,067.22 &213.0            &3205 &3129 & 17.93$\pm$0.08& 17.98$\pm$0.10 &17.87$\pm$0.09 & 17.62$\pm$0.09 & 17.86$\pm$0.08 & 17.11$\pm$0.09\\
60\,072.58 &218.3            &1975 &1910 & 18.05$\pm$0.09& 18.43$\pm$0.20 &17.91$\pm$0.10 & 17.68$\pm$0.09 & 17.83$\pm$0.08 & 17.15$\pm$0.15\\
60\,077.14 &222.9            &2999 &2925 & 18.01$\pm$0.07& 18.09$\pm$0.10 &17.95$\pm$0.10 & 17.59$\pm$0.08 & 17.88$\pm$0.07 & 16.94$\pm$0.08 \\
60\,082.04 & 227.8           &2704 &2655 & 18.09$\pm$0.07 & 18.20$\pm$0.11 & 17.86$\pm$0.10 & 17.73$\pm$0.09 & 17.82$\pm$0.07 & 16.94$\pm$0.09\\
60\,087.00& 232.7            &2446 &2375 & 18.20$\pm$0.13& 18.13$\pm$0.17&17.82$\pm$0.15&17.55$\pm$0.14&17.82$\pm$0.12&17.44$\pm$0.20\\
\hline
SED FIT & host & - & - & 18.41$\pm$0.10& 18.53$\pm$0.10&18.07$\pm$0.10& 17.70$\pm$0.10&17.48$\pm$0.10 &16.70$\pm$0.10\\
\hline
\end{tabular}
\end{center}
Notes: (1) MJD date of observations; (2) Phase (days) with respect to the discovery date MJD 59\,853.28; (3) XRT exposure time; (4) UVOT exposure time; from (5) to (10) apparent magnitudes and uncertainties for the UVOT filters \textit {UVW2}, \textit{UVM2}, \textit {UVW1}, \textit {U}, \textit {B} and \textit {V} reported in Vega system. In the last line we report the synthetic host UVOT magnitudes (in Vega system) obtained from the AT\,2022wtn host galaxy SED fitting.
All the magnitudes reported are uncorrected for foreground extinction.
With  $\cdots$ we indicate epochs with no data available (no observations). 
The XRT observation have been obtained in photon counting mode.
\noindent
\end{minipage}
\label{tbl:UVOTphot}
\end{table*} 

\bsp	
\label{lastpage}
\end{document}